\documentclass[11pt]{article}

\usepackage{scicite}

\usepackage{graphicx} 
\usepackage{times}
\usepackage{caption}
\usepackage{comment}

\usepackage{mathptmx}

\newenvironment{sciabstract}{%
\begin{quote} \bf}
{\end{quote}}

\newcounter{lastnote}

\def\nat{Nature}
\def\apj{ApJ}
\def\aap{A\&A}
\def\mnras{MNRAS}
\def\pasp{PASP}

\def\aj{AJ}

\def\apjl{ApJL}
\def\nar{NewAstRev}
\def\sovast{Soviet Ast.}
\def\apss{APSS}
\usepackage{color}

\title{Identification of the Long-Sought Common-Envelope Events.}

\author
{N.\ Ivanova$^{1\ast}$, S.\ Justham$^2$, J.L.\ Avendano Nandez$^1$  \&  J.C.\ Lombardi Jr.$^3$\\
\\
\normalsize{$^{1}$Dept.\ of Physics, University of Alberta, Edmonton, AB, T6G 2E7, Canada,}\\
\normalsize{$^{2}$National Astronomical Observatories, The Chinese
  Academy of Sciences, Beijing 100012, China,}\\
\normalsize{$^{3}$Department of Physics, Allegheny College, Meadville, PA 16335, USA}\\
\\
\normalsize{$^\ast$To whom correspondence should be addressed. E-mail:  nata.ivanova@ualberta.ca.}
}

\date{}

\begin{document} 



\baselineskip16pt


\maketitle


\begin{sciabstract}
Common-envelope events (CEEs), during which two stars temporarily orbit within a shared envelope, 
are believed to be vital for the formation of a wide range of close
binaries.  For decades, the only evidence that CEEs actually occur has
been indirect, based on the existence of systems that 
could not be otherwise explained.  Here we propose a direct
observational signature of CEE arising from a physical model where
emission from matter ejected in a CEE is controlled by a recombination front as the
matter cools. The natural range of timescales and energies
from this model, as well the expected colors, light-curve shapes, ejection velocities and event rate, match those of
a recently-recognized class of red transient outbursts.
\end{sciabstract}

Many binary star systems, including X-ray binaries, cataclysmic variables, close double-neutron stars, and the potential progenitors of Type Ia supernovae and short-duration $\gamma$-ray bursts, are thought to be formed by CEEs. 
Because  most stellar-mass binary merger
sources for gravitational waves have experienced a CEE in their
past, improved knowledge of CEEs should decrease the large uncertainty in 
theoretically-predicted merger rates. 
However, the short timescale expected for CEEs suggested that we would never directly observe them, 
allowing us only to draw inferences from the systems produced.

A CEE begins when a binary orbit becomes unstable and
decays. This might, for example, be driven purely by tidal forces
(i.e.\ the Darwin instability), although CEEs are more commonly imagined as
following a period of rapid mass transfer from one star to the
other \cite{Ivanova11}. 
In some cases the rate of transfer is so high that the
receiving star is unable to accrete all the matter without forming a
shared common envelope (CE) around the binary. This CE causes drag
on one or both stars  and hence orbital decay,
with orbital energy and angular momentum being transferred to the CE. 
This may end with a stellar merger or -- if the CE is ejected -- the binary
can survive, typically with a much reduced orbital separation,
critical to explaining many observed compact binaries.

When a CEE results in formation of a close binary, it is expected that a
substantial proportion of the mass is ejected -- typically almost the
entire envelope of one of the stars.  
Some mass can also be ejected in the case of a merger.
This partial ejection has two causes \cite{som}. 
Firstly, the orbital energy deposited into the CE early in the merger
may exceed the binding energy of the outer layers. Secondly, the upper layers of the
envelope absorb a substantial amount of the orbital angular momentum
of the companion, and angular momentum transport may be too slow to
share this across the envelope as a whole.

Here we consider the behavior of this ejected matter to try
to predict the appearance of CEEs. 
A situation involving similar physics -- type IIP supernovae
-- has been previously studied (e.g,
\cite{1991SvAL...17..210C,Popov1993,Kasen09}). 
In that model, as the ejected stellar plasma expands and cools, recombination 
changes its opacity, leading to the propagation of
a photosphere-defining ``cooling wave,'' which moves inwards with
respect to the mass variable.   

For smooth and spherically-symmetric ejecta distributions, 
the model light-curve will have a plateau shape: 
the area of the photosphere is defined by recombination, and
so the emitting surface does not grow with the speed at which the
ejected matter itself moves.
During this phase, whilst material ejected by
  the CEE will expand with velocity of the order of magnitude of the
  initial escape velocity, the photospheric radius should appear
  almost constant.
The luminosity $L_{\rm P}$ of the 
emission during the plateau \cite{1991SvAL...17..210C,Popov1993,Kasen09},
re-scaled to the likely energy range of CEE, is

\begin{equation}
\label{eqLP}
L_{\rm P}  \approx   1.7\times 10^4  L_\odot  \left(\frac {R_{\rm init} }{ 3.5 R_\odot}\right)^{2/3}    
\left ( \frac{E_{\rm k}^\infty}{10^{46} {\rm erg}}\right )^{5/6}    
\left ( \frac{m_{\rm unb}}{ 0.03 M_\odot} \right)^{-1/2}  
\left ( \frac{\kappa}{0.32  {\rm cm^{2} g^{-1}}}\right )^{-1/3}    
\left ( \frac{T_{\rm rec}}{4500 {\rm K}} \right)^{4/3}
\end{equation}
where $R_{\rm init}$ is the initial radius, $E_{\rm k}^\infty$ is the kinetic energy that the unbound mass
$m_{\rm unb}$ has at late times after escaping the potential well, 
$\kappa$  is the opacity of the ionized ejecta,
and $T_{\rm rec}$ is the recombination temperature. 
The duration of the plateau $t_{\rm  P}$ with the same assumptions is

\begin{equation}
t_{\rm P}  \approx   17\ {\rm days}  \left(\frac {R_{\rm init} }{ 3.5 R_\odot}\right)^{1/6}        
\left ( \frac{E_{\rm k}^\infty}{ 10^{46} {\rm erg} }\right )^{-1/6}    
\left ( \frac{ m_{\rm unb}}{ 0.03 M_\odot} \right)^{1/2}  
\left ( \frac{\kappa}{0.32  {\rm cm^{2} g^{-1}}}\right )^{1/6}    
\left ( \frac{T_{\rm rec}}{4500 {\rm K}} \right)^{-2/3}   ~.
\label{eqtP}
\end{equation}

This model does not depend on the origin of the
energy released during the outburst.  For type IIP supernovae, recombination controls
the release of the internal energy generated by strong supernova shocks.
For CEEs, however, there is no such supernova-provided energy input. Instead
the energy released by recombination itself may dominate
the energy budget of many outbursts \cite{2010ApJ...714..155K}. 
The unbound mass $m_{\rm unb}$ could potentially
radiate -- simply due to recombination -- as much energy as
\begin{equation}
E_{\rm recom}
 \simeq   2.6 \times 10^{46}{\rm \ ergs \ } \left ( X  + 
1.5 Y f_{\rm He} \right ) \frac{  m_{\rm unb}}{M_\odot} \ .
\label{recomb}
\end{equation}
Here $X$ is the mass fraction of hydrogen and $Y$ is the mass fraction of helium.
Hydrogen would initially be ionized in almost all of the likely ejected material from most stars; 
however helium may be fully ionized only in some fraction of it,
denoted $f_{\rm He}$.
The role of recombination in a CEE has hitherto been a debated issue 
in the overall energy balance, the controversy arising from whether it 
can be effectively converted into mechanical energy to help eject the CE \cite{1993PASP..105.1373I,Han+1994,Web08}.
This energy budget for the outburst may be increased by the thermal energy of the
ejecta. Much of the pre-CEE thermal energy of the ejecta may be expended on adiabatic cooling
\cite{2010ApJ...714..155K}. However, the shock-heating caused by the
CEE could well be substantial in some cases.

We now estimate the extent of the parameter space of CEE outbursts,
using the model described above to predict the diversity of real events. We assume that
$E_{\rm k}^\infty$  scales with the gravitational potential at the surface of
the primary star \cite{som}, and use the dimensionless
factor $\zeta$ to write $E_{\rm k}^\infty = \zeta  (G m_1^{2} f_{\rm m})/R_{\rm init}$, 
where $f_{\rm m}=m_{\rm unb}/m_1$ is the
fraction of the total primary mass $m_1$ that becomes unbound.
From equations \ref{eqLP} and \ref{eqtP}, this leads to
$L_{\rm P} \propto \left( f_{\rm m}^{2} m_1^{7}  R_{\rm init}^{-1}  \zeta^{5} \right)^{1/6}$
and 
$t_{\rm P} \propto \left( f_{\rm m}^{2}  m_1  R_{\rm init}^{2} \zeta^{-1} \right)^{1/6}$.
Two families of events seem likely, one for mergers (i.e. $f_{\rm m}
\ll 1$), and one for CE ejection (i.e. $f_{\rm m} \le 1$) (Fig.~1).

\begin{figure}
\includegraphics[width=80mm]{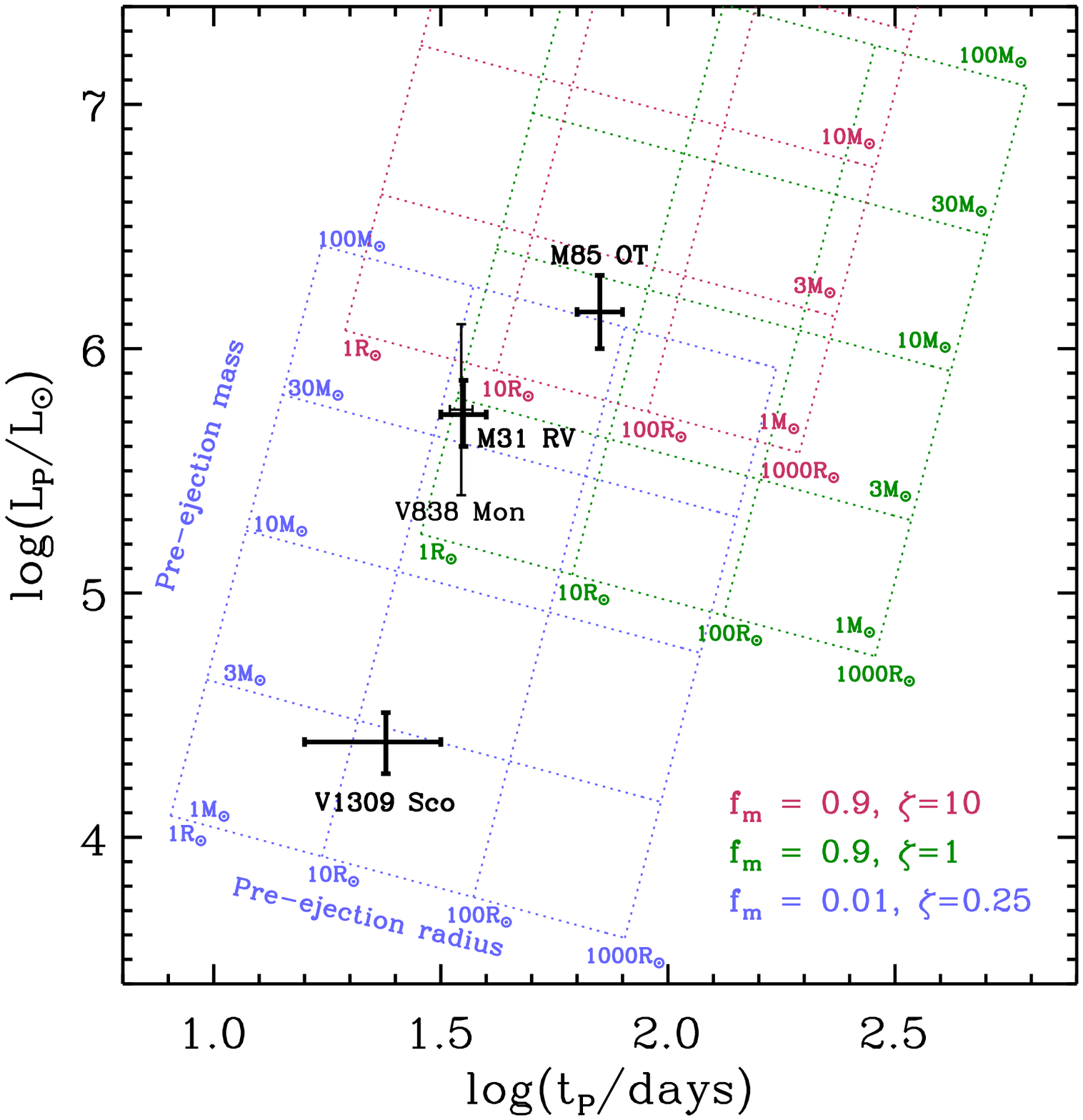}
\includegraphics[width=80mm]{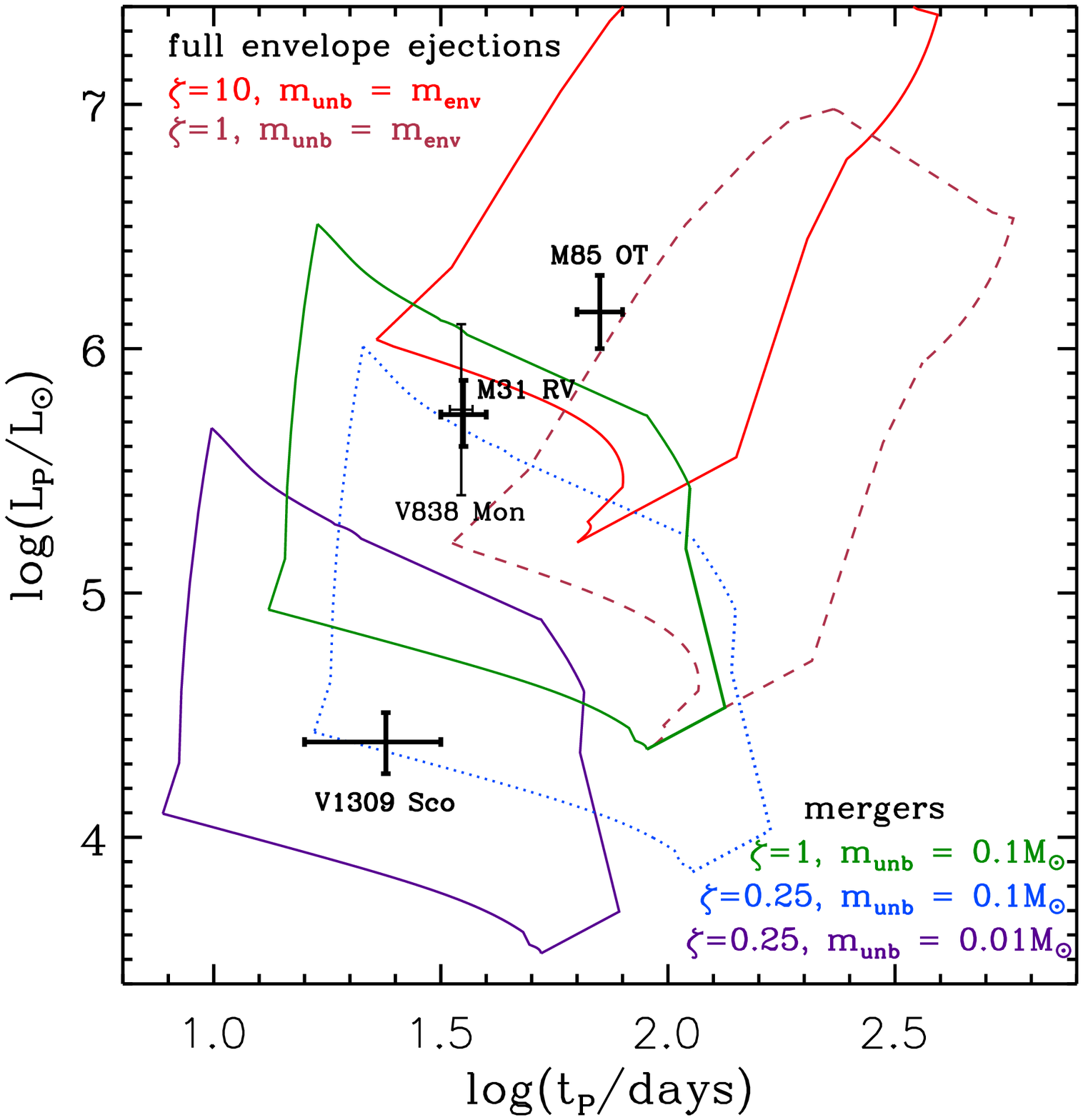}
\caption{
{\bf Left panel:} Model diversity in the $L_{\rm P}-t_{\rm P}$
parameter space is indicated by lines representing constant
  primary mass and radius. \ $f_{\rm m}$ is 
  fractional mass loss and $\zeta$ 
is the kinetic energy at infinity, parametrized as a fraction
  of the binding energy at the surface of the primary star.
Stellar mergers are in a regime of little mass ejection, while $f_{\rm m}=0.9$ approximates full envelope ejection.
{\bf Right panel:} Estimated ranges of the plateau luminosity $L_{\rm P}$ and duration $t_{\rm P}$ for
  primary stars with ZAMS masses from 1 to 150 $M_{\odot}$. 
$m_{\rm unb}$ is the ejecta mass.
  It is assumed that mergers can happen anytime during the primary's evolution,
 whereas full envelope ejection
  can occur only for post-MS primary stars. 
  We used  fitting formulae for stellar evolution  \cite{Hurley+2000},
  at $\rm Z_{\odot}$.  In both panels, values for $L_{\rm P}$ and  $t_{\rm P}$
 are marked for the outbursts from V1309 Sco, M85 OT, M31-RV, and V838 Mon.
}
\label{fig:grids}
\end{figure}

In addition to the predicted ranges of outburst energy and duration,
this model for CEE outbursts has several noteworthy features. 
The physics that causes the plateau-shaped light-curve should lead to
a difference in the phometrically-inferred expansion velocity and the
actual material velocity (which could be inferred from spectra). The 
effective photospheric temperature should be $\sim 5000$K  for thick ejecta 
\cite{Popov1993}, and so the outburst color will naturally be red. 
In addition, once the ejected envelope has fully recombined, 
the material may suddenly become transparent --- unless enough of the ejecta has cooled down sufficiently to produce dust.
These characteristics are reminiscent of curious transients with
predominantly red spectra recently detected in the local universe 
(e.g., \cite{1990ApJ...353L..35M,1999AJ....118.1034M,Bond+2003,2007Natur.447..458K,
2007Natur.449E...1P,Bond+2009,2011ApJ...737...17B,Kasliwal+2011})).  
This empirical class has been dubbed Luminous Red Novae (LRNe),
a subset of the even more ambiguously defined class of 
intermediate luminosity red transients (ILRTs) \cite{som}. 
ILRTs cover a wide range
of outburst energies -- from  $10^{45}$  to a few $10^{47}$
ergs (brighter than the brightest novae, but still
fainter than Type Ia supernovae). 
They are characterized by spectroscopically inferred expansion velocities of 200-1000 km/s  -- much lower than would be
  expected for novae or supernovae, and also strikingly different from the
 photometric expansion velocities
\cite{2009ApJ...699.1850B}.
In addition, some could be seen as red giants within a dozen years after the outburst \cite{2011ApJ...737...17B, Tylenda11}.

It was not known what ILRTs are or whether they have a common cause;
several ideas have been suggested \cite{som}.
A model which considered the possibility that
LRNe are caused by stellar mergers -- a subset of CEEs -- has been
independently considered several times for different LRN outbursts,
though further examinations of outburst features always showed various drawbacks.
However, those problematic features do match expectations from our CEE-driven outburst model \cite{som}.

A particular feature of the LRN outbursts -- as opposed to
all ILRTs -- is the presence of a plateau in their luminosity
curves. We compare well-known LRNe  \cite{som} to the expected CEE
diversity in Fig.~1.
The agreement is striking, especially given the simplicity of the model and the
potential complexities it neglects -- e.g., how CEE ejecta deviate from 
spherical symmetry, or how much $\zeta$ for mergers might be different 
to $\zeta$ for full envelope ejection \cite{som}.

M85 OT2006-1 is a LRN with well-known peak luminosity and plateau duration. 
If the luminosity from M85 OT2006-1 was largely from recombination,  $\sim 1.5 M_\odot$ 
of plasma would have recombined to provide the observed total energy. 
This fits with constraints on the progenitor mass ($\le 2 M_\odot$) from the stellar population
age \cite{Of08}. 
Thus M85 OT2006-1 plausibly 
ejected the whole envelope of a low-mass giant.
This outburst showed a plateau, with luminosity $\approx 1$--$2\times
10^6 L_\odot$ \cite{Rau07} for $\approx$60--80 days,
and had expansion velocities of 350 km/s \cite{2007Natur.449E...1P}. 
Our inferred ejecta mass and that observed expansion velocity indicate a kinetic energy of
$\sim 1.8\times 10^{48}$ erg. Then $R_{\rm init}=45 R_\odot$, self-consistent with our model, 
gives $L_P\sim 10^6 L_\odot$ and $t_P\sim 70$ days.
 
Another recent outburst, V1309 Sco, is similar to, but fainter than, most LRNe,  as it radiated away
 only $\sim 3\times 10^{44}$ erg during a $\sim25$-day plateau-shaped maximum in the light-curve \cite{Tylenda11}. 
The progenitor was a contact binary with a relatively rapidly-decaying
orbital period of $\sim1.4$ day. 
After the outburst, the system appeared to be a single star; therefore this appears to have been a CEE, 
leading to a merger \cite{Tylenda11}.  
However, several features of the V1309 Sco outburst -- in particular the plateau in the light-curve and sudden transparency --
were difficult to reconcile with prior theoretical expectations for the appearance of a CEE \cite{som}.  

Because the V1309 Sco progenitor was observed in detail,
this system is ideal for testing our model. 
Beginning with the properties of the pre-merger contact binary \cite{Stepien11,Tylenda11}, we calculated the amount of
material that became unbound during the V1309 merger using
two methods -- simple energy balance using a 1D stellar code and a
set of 3D hydrodynamical simulations \cite{som}.
Both methods predict that a small mass, $\sim 0.03$ to $0.08 M_\odot$, will become unbound.
Complete recombination of this ejected mass would provide enough energy ($\ge 7\times 10^{44}$ ergs) 
to explain the total energy output of V1309 Sco.
The output of the 3D simulations, combined with Equations \ref{eqLP} and \ref{eqtP},
predicts plateau durations from 16 to 25 days, and plateau luminosities 
of $1.8$--$2.2\times10^4L_\odot$. These values match the
observed luminosity ($L_P\sim 2\times10^4L_\odot$) and 
plateau duration (about 25 days).

Considering the menagerie of theoretically-expected outbursts
from CEEs, we note that events in the top-right of Fig.~1 should be relatively rare 
[compare with $\eta$ Car; see \cite{som}], and those in the bottom-left (stellar mergers) 
comparatively hard to detect in a magnitude-limited survey.
Assuming that the peak luminosity of the outburst is about an order of magnitude higher
than $L_{\rm P}$, we find that the whole range of $L_{\rm P}$ and $t_{\rm P}$ for stellar masses $1-150 M_\odot$
coincides well with the observed domain for luminosities and durations
of LRNe suggested in \cite{Rau09}.
We can estimate the rate of CEE-originated outbursts that 
appear as red transients, by considering what fraction of stars in the galaxy
undergo a CEE.  
We estimate 0.024 such events per year per Milky Way-like galaxy \cite{som}, 
of which about half should be more luminous outbursts (results of a CE ejection), and half are lower-luminosity events
(powered by stellar mergers).
This is consistent with the empirical lower limit for more luminous ILRTs of 0.019 ${\rm yr^{-1}}$ for the
Galaxy \cite{Of08}, because we do not expect that all luminous ILRTs must be powered by a CEE
[though some non-LRN ILRTs-- like NGC~300-OT or SN2008S -- might potentially also be triggered by CEEs
\cite{som}].

The question of whether recombination energy can help to unbind a stellar envelope during a CEE
is important for understanding the formation and survival of many binary systems \cite{Han+1994,Web08}.
Our model suggests that a large fraction of the energy from recombination is
commonly radiated away following a CEE.
Such luminosity provides a beacon, which helps to
illuminate and identify a CE ejection or merger at large distances. 
The recombination wave also controls the shape of the plateau-shaped
light-curve of LRNe. 
We therefore suggest that detecting and characterizing the population
of ILRTs will help us understand CEEs.

\ 

\noindent{\Large \bf Supplementary text}

Here we provide additional information about our hydrodynamical simulations, as well as give more details behind our formulae and arguments from the main text.

\ 

\noindent{\large \bf S1. Intermediate Luminosity Red Transients}  

In recent years, observers have identified a new class of transients,
with peak luminosities
somewhere between that of the brightest novae and Type Ia SNe
and with a total energy output anywhere from  $10^{45}$  to a few $10^{47}$ ergs, 
e.g., \cite{1990ApJ...353L..35M,1999AJ....118.1034M,Bond+2003,2007Natur.447..458K,
2007Natur.449E...1P,Bond+2009,2011ApJ...737...17B,Kasliwal+2011}. 
Because their spectra are predominantly red, completely unlike novae, they have been called
intermediate luminosity red transients  (ILRTs, \cite{2012AAS...21943609B}).
However alternate names are also in common use, either for all of the
transients or for particular kinds of events,  e.g. Luminous Red Novae
(LRNe), intermediate luminosity optical transients,
mid infra-red transients, V838 Mon-like events, and supernova
impostors. The relative novelty of these classes, and the
  uncertainty over their physical origins, makes it unsurprising that
  the use of such terms is not necessarily always consistent between
  different papers. 

It is not known what red transients are or whether they have a common cause:
it has been argued that ILRTs could be due to
accretion-powered jets \cite{2012ApJ...746..100S}, 
tidal disruption of planets \cite{2011MNRAS.416.1965B},
nuclear outbursts \cite{1992ApJ...389..369I}, electron-capture SNe \cite{2009ApJ...705.1364T}
with shocks propagating through dusty surroundings \cite{2011ApJ...741...37K}, 
 or violent stellar mergers \cite{2006MNRAS.373..733S}.
It has also been argued that -- despite having  similar outburst properties in terms of color
and luminosity -- ILRTs may have different origins \cite{2012AAS...21943609B}.

In this work, we consider the particular group of red transients which
are most frequently labeled as LRNe.  
The widely-recognized members of this class of red transients
are M31-RV, V838 Mon, and  M85 OT2006-1.  We note that the recognition of LRNe as a 
special class of stellar explosion started with the last object in
that list, although it was not the first of them to be detected.
In this paper we also include V1309 Sco as a possible member of this
LRN class, at the low-energy end \cite{Tylenda11}.
Another suggested low-energy end LRN is V4332 Sgr -- this transient radiated away a 
total energy of only $4.5\times 10^{43}$ ergs \cite{2005A&A...439..651T}, an order of magnitude less even than V1309 Sco.
The LRN class is likely separated from another class of red transients 
where progenitors were observationally identified to be dusty
modestly-massive stars (with  $M \sim 10 M_\odot$) -- as for example in the cases of SN 2008S and NGC 300 OT 
\cite{2009ApJ...699.1850B,2008ApJ...681L...9P,2009ApJ...697L..49S}; 
those ILRTs are better known as supernova impostors.
Nonetheless, because the recognition of red transients as a new type of astrophysical 
object is very recent, their classification is not yet by any means universally established  
or finally accepted throughout the astrophysics community.

We note that some investigators have suggested that LRNe could be present only in old populations, 
with V838 Mon standing as an exception
(see also the discussion against this point of view in \cite{2007ASPC..363..189S}). 
It is important to clarify that our proposed link between CEEs  
and LRNe implies that there should not be any such restriction on population age for LRNe: 
our CEE outburst model is related only to the reason of the ejection 
-- a CEE -- and to the  self-similar physics and recombination outburst following the CEE.
Indeed, CEEs may occur in binary stars with companions of any masses. In terms of rates, 
CEE from massive stars are expected to be much more rare, as their relative fraction 
in the total number of stars is small; however they are not forbidden.
In addition, given the current understanding of CEE physics, 
it is impossible to say how the CE ejection would differ in low-mass and high-mass stars.

\ 

\noindent{\large \bf S2. Previous problems with a merger model: what does
  the standard CEE model predict? }

One of the most common models suggested to explain different LRNe -- V838 Mon, M85 OT 2006-1 and V1309 Sco --
is a merger, either of two stars, or of a star with a planet 
(e.g.,\cite{2003ApJ...582L.105S,2006MNRAS.373..733S,2007Natur.447..458K, Tylenda11}).
It is also known as the merger-burst model;
however, because a merger is a particular type of CEE, we will
hereafter refer to it as to a CEE model.
The case of V1309 Sco is rather unequivocally a CEE, as observations 
revealed a binary prior the outburst and a single star after\cite{Tylenda11}.
However, there were problems with justifying the physics behind this model.
For example, a stellar merger scenario in the case of M85 OT2006-1 was rejected  because 
an analysis of the energy available from a merger suggested that a violent merger would be unable, 
by a factor of a few, to explain the energy that was radiated away from M85 OT2006-1 \cite{Of08}. 
In \cite{2004A&A...418..869B}, striking similarities between the M31-RV, V838 Mon and V4332 Sgr outbursts 
were noticed, leading to a discussion which concluded that the merger-powered
outburst models proposed to-that-date  
showed too much dependency on metallicty, mass, and ages and hence 
could not explain the observed  homology.

We list next a number of striking observational features of LRNe which 
were not previously theoretically anticipated by a CEE model   
(one leading to either a merger or a the formation of a close binary)  
and which were not emphasized as discriminating features 
in previous studies:

\begin{itemize}

\item {\bf Large increases in radius and luminosity.} In previously published CE
  simulations, the increase in the stellar radius of the bound mass during
  fast spiral-in is usually less than a factor of 10 (see
  e.g., \cite{Passy12, Ricker12}). These simulations were performed for
  binaries that are expected to survive, rather than merge,
  i.e.\ these binaries have more energy injected into the stellar
  envelope -- relative to the binding energy -- than for binaries
  that merge.  However the well-studied example V 1309 Sco
  reached a maximum luminosity $\sim 3\times 10^4 L_\odot$, about
  $\sim 5000$ times larger than the initial luminosity of the progenitor
  ($\sim3.0-8.6 L_\odot$), whilst
its effective temperature dropped to $\sim 4000$K during the peak
luminosity and plateau \cite{Tylenda11}. 
This indicates an increase in its apparent (or effective) radius by about 90 times, reaching 
$\sim 300 R_\odot$ \cite{Tylenda11}, at least
an order of magnitude more than simulations suggested.
Similarly, interferometric observation showed that V838 Mon expanded by $\sim 300$ times during the outburst, 
to $1570\pm 400 R_\odot$, while its progenitor was likely 
$5-10 M_\odot$ with a radius of only about $5 R_\odot$ \cite{TylendaSoker05}.

 \item {\bf An unexpectedly long duration.}  Expansion of the envelope during CEE is expected to occur
during the dynamical plunge-in phase (see, e.g.,\cite{Ivanova11}),
i.e.\ the stage during which the inspiralling star loses most of its angular momentum. 
It is theoretically expected, and confirmed by numerical simulations (see e.g. \cite{Passy12, Ricker12}),
that this phase proceeds on a timescale of about several
\emph{initial} dynamical timescales of the stellar envelope of the
primary star. This is a few days for the case of V1309 Sco; however, the outburst
lasts $\approx$100 days (\cite{Tylenda11}, also see Fig. S1) 
which is comparable to dynamical timescale of this object at the
\emph{maximum} of expansion, $\tau_{\rm dyn, max} \approx 80$ days.
In the case of V838 Mon, the outburst last for about 80 days \cite{crause03} while the initial dynamical timescale
would be only a couple of hours. In case of M85 OT2006-1, in which the outburst last about 60 days \cite{2007Natur.449E...1P},
there are no strong constraints on progenitor.

\item{\bf Plateau phase.} Several ILRTs outbursts, after the initial rise in brightness, 
featured a plateau in their light curves in the red band at a luminosity somewhat lower than at the maximum.  
Specifically, in case of V1309 Sco the plateau duration is $\sim 25$ days (depending on the 
definition of the start and the end of the plateau phase,  from 16 to 31 days (\cite{Tylenda11}), also see Fig. S1),
$\sim 60-80$ days in case of M85 OT2006-1 \cite{2007Natur.449E...1P}, $\sim 15$ days in case of V4332 Sgr 
(\cite{2005A&A...439..651T}, although the start of the outburst is not well known)
and $\sim30-40$ days in case of M31-RV \cite{2004A&A...418..869B}.
V838 Mon has a very complex visual band light-curve with several peaks
separated by about 30 days.  Nonethless, it does have an easily
distinguished plateau in the red part of the spectrum as well as in
bolometric luminosity; the overall duration of the outburst is about
80 days, and that plateau starts $\sim 30$ 
days after the peak luminosity and lasts for another $\sim 35$ days \cite{2003MNRAS.345L..25R, 2005A&A...436.1009T}.
Most of the energy in LRNe is radiated away during the plateau
phase. 
The plateau shape had not been predicted and had not yet been explained.

 \item {\bf An extremely rapid decline.} The post-common-envelope decline in luminosity 
of the merger product is expected to occur on a thermal timescale,
which -- for these over-luminous and over-inflated objects -- is typically about a year. 
However, the abrupt decline in luminosity (by a factor of 100) in V1309 Sco, M31-RV, V4332 Sgr and V838 Mon 
happened in a dozen days (\cite{Tylenda11,2004A&A...418..869B,2005A&A...439..651T,2003MNRAS.345L..25R, 2005A&A...436.1009T}, also see Fig. S1).
For V 1309 Sco, the timescale for exponential decrease was determined to be only a few days \cite{Tylenda11}.
Strikingly, the timescale for this sharp luminosity decrease   
is considerably shorter than the dynamical timescale $\tau_{\rm dyn, max}$ at the maximum of expansion for V 1309 Sco (80 days)
or for V838 Mon (about 500 days).

\item{\bf Inconsistent velocities.} While expansion velocities from
the spectra were several hundred km/s (e.g., in V1309 Sco \cite{Tylenda11},  in M85 OT2006-1 \cite{2007Natur.449E...1P}, and in V838Mon \cite{2003ApJ...588..486W}), 
velocities inferred from the apparent radius expansion imply speeds from only about 20 km/s (V 1309 Sco) to about 100 km/s (V838 Mon).

\end{itemize}

In section S3 below we explain how the model presented in this
  paper naturally accounts for the above features.

\

\noindent{\large \bf S3. Wavefront of cooling and recombination in a CEE}

As discussed above, the light curves of LRNs have five 
similar striking features, where two are the most exceptional in providing important clues to the physics of the outburst.
First, the outbursts have a roughly constant plateau luminosity $L_{\rm P}$ for a time $t_{\rm P}$ that is typically dozens of days (for a note on the special case of V838 Mon see, see \S S5).
Second, the inferred radius of the photosphere increases
relatively slowly during most of the plateau phase, at the end of which the apparent radius  
decreases very quickly.  This is reminiscent of the behaviour of type IIP
supernovae, in which the photosphere does not stay at a fixed
Lagrangian coordinate but moves inwards in mass as the ejecta expands
and cools; roughly self-similar, homologous expansion means that the radius of the
photosphere and luminosity of the emission remain approximately
constant \cite{ImsNad1965, Grassberg+1971, GraNad1976, Popov1993,
  Eastman+1994}.

Such a photosphere-defining `cooling wave,' propagating inward in the frame of
the expanding shock, was predicted for terrestrial explosions in air by \cite{ZKR1958a,ZKR1958b}. 
This type of behaviour does not need to be
associated with recombination; it is only dependent on a non-linear
temperature dependence of the opacity of the ejecta (see also
\cite{Bethe1964}, 
which argues that the necessary condition is that the effective opacity
increases more rapidly than a cubic power of temperature).  
The low-opacity material outside the photosphere
is able to cool by radiation much more effectively than the hot,
high-opacity material inside.
However, in the case considered here, as well as for type IIP SNe,
recombination is the cause of the change in opacity that leads to
this cooling-wave. The location of the effective photosphere is
controlled by a sudden large reduction in the mean opacity of the 
ejecta after hydrogen recombination.

Even though the mean opacity greatly drops outside the region of
recombination, the photons emitted during recombination will not in general escape directly.
Although neutral hydrogen has a much lower Rosseland mean
opacity when compared to ionised hydrogen, the relevant line
opacity of neutral hydrogen is higher than for the ionised
material. So the recombination photons themselves have a short mean
free path, and there is no reason to expect strong $\rm H_{\alpha}$ line
emission, even if photons at other wavelengths are free to escape. 
Uniformly applying the Rosseland mean opacity in this case
would be misleading, and full wavelength-dependent radiative transfer
would be necessary to simulate precisely the structure of the cooling
wave.  Hence situations where the location of the cooling wave is
defined by recombination are more complicated than the cases
first considered by \cite{ZKR1958a,ZKR1958b}, in particular because there is a
release of energy from recombination that is coincident with the
change in opacity.  However, we consider that
the energy input is so close to the photosphere that it
justifies neglecting the effect of any heating on the main
features of the hydrodynamics. 

This physics is well-studied in the context of supernovae, and we can
apply appropriate results here.  The analytic model of \cite{Popov1993},
based on the idea of there being a wavefront of cooling and recombination (WCR),
is particularly attractive.  Although the method produces only approximate light curves, it
provides accurate scaling relations that give insight into the underlying
physics and to the interpretation of observations.  Indeed, \cite{Kasen09}
verifies the major results of the Popov model with detailed hydrodynamical 
simulations that include radiative transfer.

The analytic WCR model has the following intrinsic features:
\begin{enumerate}
\item[a)] An outburst has a plateau phase. 
(For more on the condition which must be satisfied in order to obtain a plateau, see below.);
\item[b)] Because the observable surface during the outburst is not the stellar surface, 
but instead the photosphere of the expanding ejected material defined by the location of the WCR, a large apparent radius increase is expected;
\item[c)] As this recombination front propagates inwards through the material by mass,  its geometrical 
distance to the center can gradually increase and then decrease until the recombination of the material is over  
(see Eq.~(16) in \cite{Popov1993}), while at the same time the outflow passing through the WCR continues to stream outward. This leads to what could 
be observationally identified as an inconsistency between measured material expansion
velocities and apparent expansion of the radiating surface: the photometric velocity indicating the expansion rate of the photosphere can be much less than spectroscopic velocities, which give the speed of the gas passing through the photosphere;
\item[d)] The model connects the outburst not with the dynamical timescale on which the CE rapidly expands,
but with the timescale over which recombination wave proceeds;
\item[e)] After the recombination is completed, the model allows sudden transparency, which
could be observed as  rapid decline of luminosity and appearance of the central object. 
If material above the front is dense and cool enough, it could form
dust that would hide the central object again {(for a discussion of
dust formation in the ejecta of a related class of transients then
see, e.g., \cite{Kochanek+2012})}.
\end{enumerate}

The recombination process depends strongly on how not only
temperature, but also density, evolves.  Hydrodynamical simulations
show that recombination temperature in this model is $\sim 5000-6000$K
for thick and dense  ejected envelopes, as in case of SNIIP
\cite{Popov1993, Kasen09}.  
Somewhat colder recombination temperatures are expected in less dense ejecta, where the number density of free electrons (which are required for recombination) is smaller.  In the grey atmosphere approximation, the effective temperature is a factor of $2^{1/4}\approx 1.2$ larger than the recombination temperature, corresponding to effective temperatures in the $\sim 4000-6000$K range (see also \cite{1985SvAL...11..145L} 
who showed that effective temperatures shift to lower values once the energy of the outburst is decreasing).  Although full 3D
hydrodynamical simulations with radiative transfer are necessary for more detailed calculations of 
the effective temperature, it is clear that the WCR model predicts that a CEE outburst will be bright in the red part of the spectrum. {The above is a simplification of the complicated situation occurring in these photospheres (e.g. it neglects the role of $H^-$ ions formed due to electrons donated from metals), but the outcome of a detailed analysis should yield similar conclusions.}

For this study, we take the WCR model derived by
\cite{Popov1993}, in the convenient form expressed by
\cite{Eastman+1994}, and re-scale it to the situation we are
studying.\footnote{For equation (1) in the main text, 
we have corrected a small numerical
  error in the coefficient of equation (9) of \cite{Eastman+1994}, for the plateau luminosity, such that it is
  consistent with the \cite{Popov1993} value.}
We scaled opacities to 0.32 $\rm cm^{2}\,g^{-1}$ for $X=0.7$.
The original derivation by \cite{Popov1993} assumed that
the kinetic energy input produces self-similar expansion of an envelope moving with constant
velocity. This implies that the potential well of the exploding star
has already been overcome, and hence that energy input for his equations in the case of CEE is the
same quantity as $E_{\rm k}^\infty$.

An important assumption in Popov's model is that recombination,
although controlling the energy release  through its effects on opacity, 
does not dominate the dynamics of the expansion.
In the case of CEE, we can estimate from Eq.~(3) of the main paper
how much energy is available from recombination.  
For typical cases of partial envelope ejection, such as in the simulated case of V1309 Sco below,
the kinetic energy in the outburst is $E_{\rm k}^\infty \sim 3$--$4\times 10^{47}{\rm erg}\, m_{\rm unb}/M_\odot$, 
exceeding the recombination energy by more than an order of magnitude.
For cases  of complete CE ejection, we can estimate the kinetic energy $E_{\rm k}^\infty$ to be of the same order
as the initial potential energy of the CEE progenitor-star: $E_{\rm k}^\infty=G m_{\rm unb} (m_1 - m_{\rm unb})  (\lambda R_{\rm init})^{-1}$, 
where $\lambda$ is an envelope structure parameter that is roughly of order unity for small low-mass giants
but that can be as small as 0.01 for massive and larger giants \cite{2011ApJ...743...49L}.  Then
$$
\frac{E_{\rm k}^{\infty}}{E_{\rm recomb}} \sim 150 \times (X+1.5Y f_{\rm He})^{-1}\, \frac{m_{1}-m_{\rm unb}}{M_\odot} 
\left (\frac{\lambda R_{\rm init}}{R_\odot} \right ) ^{-1}. 
$$
So $E_{\rm k}^\infty$ easily exceeds $E_{\rm recomb}$ by at least an order of
magnitude.  Hence, although the recombination energy is a larger fraction of $E_k^\infty$ in the CEE
problem than it is in the Type IIP SNe problem, the recombination
energy is still relatively small enough that we feel justified in
assuming that it should not play a significant role in driving the
dynamics, therefore making Popov's model applicable for CEE. 

{We briefly clarify here that the analysis for $E_{\rm k}^\infty$ above must self-consistently extend to the choice of $\zeta$ for complete envelope ejection. For that reason, a typical $\zeta$ for complete envelope ejection might be expected to be somewhat higher than a typical  $\zeta$ for ejection resulting from a merger event. In those extreme cases where $\lambda = 0.01$ then an assumption of $\zeta = 100$ would become reasonable, though the range $1 < \zeta < 10$ seems likely to be more typical.}

In Popov's simplified analytical model for the WCR, the main role of recombination is
to provide the reason that the gas switches from opaque to transparent,
resulting in a plateau-shaped light-curve.  The Popov model predicts that light curves will be self-similar to each other, with their shape characterized by the dimensionless parameter $\Lambda$: larger values of $\Lambda$ correspond to more pronounced plateaus.
We find that, scaling for the case of CEE-outbursts, $\Lambda$ from \cite{Popov1993} can be calculated as

$$
\Lambda  \approx   49   \left(\frac {R_{\rm init} }{ 3.5 R_\odot}\right)^{-1}    
\left ( \frac{E_{\rm k}^\infty}{10^{46} {\rm erg}}\right )^{-1/2}    
\left ( \frac{m_{\rm unb}}{ 0.03 M_\odot} \right)^{3/2}  
\left ( \frac{\kappa}{0.32  {\rm cm^{2} g^{-1}}}\right )^{2}    
\left ( \frac{T_{\rm rec}}{4500 {\rm K}} \right)^{4},  
$$
which for cases of complete envelope ejection becomes
$$
\Lambda \approx 1700 \lambda^{1/2}
\left(\frac {R_{\rm init} }{  R_\odot}\right)^{-1/2}    
\left ( \frac{m_{\rm unb}^2}{ (m_1-m_{\rm unb})  M_\odot} \right)^{1/2}  
\left ( \frac{\kappa}{0.32  {\rm cm^{2} g^{-1}}}\right )^{2}    
\left ( \frac{T_{\rm rec}}{4500 {\rm K}} \right)^{4} .  
$$

Consequently, for most of the range of possible CEE-progenitors, $\Lambda\gg 1$, and hence a plateau phase with sudden transparency is predicted. If $\Lambda \ll 1$, which is not a likely case for CE progenitors, the model predicts slow decline with no well defined plateau in the light-curve.  Cases with intermediate values of $\Lambda\sim 1$ may provide a phase somewhat like a plateau, with the luminosity taking a comparatively long time to decline after recombination is completed (see also Fig. S2).

We note that these scaling relations were derived
under the approximation that the ejection is spherically symmetric, which may well be less accurate
here than for supernovae. In addition, \cite{Popov1993} assumed that
the plasma was radiation dominated, which may well be less
appropriate for matter ejected from a stellar merger than for supernova
ejecta. However, it has previously
been noted that at the location of the cooling wave recombination
itself will automatically lead to radiation becoming dominant
(e.g., \cite{1991SvAL...17..210C}), which helps to justify our
adoption of the Popov expressions.  
Indeed, \cite{2010ApJ...714..155K} show that during the recombination phase
the third generalized adiabatic index ($\Gamma_3$) decreases
below the normal value for radiation-pressure dominated matter,
even for initially gas-pressure dominated matter.  This allows
continued expansion by an order of magnitude or more at near-constant
temperature -- independent of whether the initial matter is
gas- or radiation-pressure dominated -- and also increases the
relative importance of radiation pressure, even in initially cold envelopes.

\

\noindent{\large \bf S4. Simulations of V1309 Sco}

For the properties of the pre-merger contact binary in V1309 Sco, we take a
low-mass subgiant with mass $M_{\rm 1} \sim 1.52 M_\odot$ and  
radius $R_{\rm 1}=3.5 R_\odot$ with a lower-mass companion of
$M_2\sim 0.16 M_\odot$ \cite{Stepien11},  with an
orbital period at Roche Lobe overflow of $P_{\rm orb}\sim
1.42$ days (the last detected orbital period before the merger, \cite{Tylenda11}).

The binding energy of the envelope of a subgiant representing V1309 Sco primary is $|E_{\rm  bind}|\simeq 1.5\times 10^{48}$ 
ergs (as is standard, here we include both the
gravitational potential and internal thermal energy terms but not the recombination term).
Stellar model calculations were performed using the {\tt STARS/ev}
stellar evolution code, originally developed by Eggleton 
\cite{Egg71, Egg72, Egg73, Eggetal73}, with recent updates described in 
\cite{Pols95,Glebbeek2008} and references therein.
For this companion mass of $0.16 M_\odot$, even if the companion 
survives the entire spiral-in and the orbital 
energy is released as close as possible to the boundary of the
primary's core (i.e.\ providing the maximum possible energy output), 
the merger would produce only about $\Delta E_{\rm orb} \simeq 5\times 10^{47}$ erg. This is only a third 
of the envelope's binding energy. So a CEE in this system necessarily
results in a merger before the envelope is unbound.

However, for this particular primary, soon after 
the companion starts to plunge-in, 
the released orbital energy 
near the surface exceeds the binding energy of that part of envelope
already outside the location of the companion. 
From the detailed stellar model, we find that the mass that can become unbound during this 
initial phase of the spiral-in is $ m_{\rm unb}\approx 0.04 M_\odot$. 
The total orbital energy that was deposited in this mass is $\sim 6 \times 10^{46}$ ergs; 
some of this material will be ejected only barely faster than the local escape velocity ($v_{\rm esc}\approx 420$km/s for the unperturbed star) and some will 
get significantly more specific kinetic energy. Ejected matter that
is given more  energy than is required for overcoming the potential barrier will
still have non-zero kinetic energy at infinity $E_{\rm k}^\infty>0$;
this will be smaller than the total energy deposited, but should be of the same order.

The magnitude of the initial velocity $v$ of the ejected upper layers can be qualitatively understood by considering a circular binary consisting of two companions with masses $m_1$ and $m_2$, orbital separation $a$,
 and total orbital angular momentum $J$ approximated by Keplerian two-body expressions: $J=\mu\Omega a^2=m_1m_2(Ga/M)^{1/2}$, where the reduced mass $\mu=m_1m_2/M$, $\Omega^2=GM/a^3$, and $M=m_1+m_2$.
To shrink the orbit by $da$, 
a fraction of the orbital angular momentum $dJ$ must be transferred to a portion $dm_1$ of the stellar envelope material.
This angular momentum transfer occurs in the very top layers of the
primary star, and hence the stellar mass which is inside the orbit of the
companion star does not significantly change.  Therefore, treating $m_1$ and $m_2$ as constant, 
$$
dJ\approx \frac{1}{2} \frac{J}{a} da=\frac{1}{2}\mu\Omega a da\ .
$$

\noindent When the companion already orbits inside the envelope,  $a\approx r_1$ (although the $r_1$ describing
particular particles of matter can be somewhat larger than their initial $r_1$ in an unperturbed star).  In the process,
the stellar material $d m_1$ achieves an angular momentum $dJ= v r_1 dm_1$, implying a tangential velocity
$$
v \approx \frac{1}{a}\frac{dJ}{dm_1}  \approx  \frac{1}{2} \mu\Omega 
\left ( \frac{dm_1}{dr_1}\right )^{-1} .
$$

For the masses and initial orbital period used in our V1309~Sco simulations, we find that this initial tangential velocity is
$$
v\approx 2.5 {\rm \ km\ s}^{-1}\  \left ( \frac{dm_1/M_\odot}{dr_1/R_\odot} \right )^{-1}.
$$
From the mass profile of the larger star's envelope, 
$dm_{\rm 1}/dr_1=4\pi r_1^2\rho(r_1) $, where $\rho(r_1)$ is the density profile and $r_1$ is the distance to the center of the donor. 
In the upper $0.02 M_\odot$ of our pre-merger 1.52 $M_\odot$ star
$(dm_1/M_\odot)/ (dr_{1}/R_\odot) \sim 10^{-4}$--$10^{-1}$,  with smaller values closer to the surface. 
We note that when mass ejection begins, the layers will expand and somewhat more mass 
will have even smaller values of $(dm_1/M_\odot)/(dr_{\rm 1}/R_\odot)$.
A significant fraction of this upper envelope therefore is expected to be ejected at 
velocities higher than $v_{\rm esc}$. 
Considering the pre-merger orbital configuration, the expected
kinetic energy at infinity is of order  $\sim 10^{46}$ ergs.

The estimates above were made for a symmetric, non-rotating 1D stellar
model. For a real -- 3D -- star filling its Roche lobe, most of the
companion-envelope interactions occur close to the orbital
plane. There the star's radius would be larger than for 1D models, so
we expect that more mass could be lost in the rotating case. Yet, since most of this expansion
of the stellar structure happens in the very outer layers, we do not expect
the enhancement to be more than a factor of a few.

From the stellar model, we also find that  $f_{\rm He}\approx 0.8$. 
Hence,  if all mass that could become unbound would also recombine, 
the total energy that can be radiated away during the recombination 
in V1309 Sco case is $\sim 10^{45}$ ergs -- so
the available recombination energy reservoir can explain the observed
outburst's energy very well.

For a better estimate about how much mass could be ejected,
we performed several 
numerical simulations using the 3D SPH code {\tt StarCrash} \cite{Lombardi06,Gaburov10}. 
This code was specifically re-developed to deal with close binary systems \cite{Lombardi11}.
For these numerical studies, we varied the initial orbital period
around the observed pre-merger value $P_{\rm orb}\sim 1.42$ day
\cite{Tylenda11}. Table~S1 gives the complete list of initial
conditions, including rotational synchronization of the giant. The
companion was modelled as either a low-mass main sequence star (with
SPH particles) or as a point mass (representing the core of a red
giant which lost its envelope in a previous binary interaction
\cite{Stepien11}). Stellar structures were first calculated using the
{\tt STARS/ev} code and then relaxed in  {\tt StarCrash} in a binary
configuration close to Roche-lobe overflow. As a result of this
relaxation, the radius of the giant in the orbital plane was slightly
larger than its 1D radius obtained with the stellar code, as expected
\cite{Renvoize+2002}.

A visualization of simulation ps334 presents the evolution of the
column density (in g/cm$^2$) as viewed from a direction perpendicular
to the orbital plane (see Movie S1).  After the merger, $\sim 0.06 M_\odot$ of unbound
material is left streaming from the merger product with some deviation
from axisymmetry, including clumpiness. 
In all simulations the merger ejects a small fraction of the donor, and the unbound mass is comparable with our preliminary estimates, $\sim 0.04 M_\odot$ (depending on the degree of corotation, it varies from $\sim 0.03$ to $0.08 M_\odot$, see Table~S1). This mass is obtained by computing the total energy (kinetic, gravitational and internal) for each particle: if this energy is positive, the particle is considered to be unbound. We find that for the same donors, having a MS companion will result in less mass ejection. For most simulations, the mass is ejected in two episodes (see Fig.~S1), where during the first episode the companion is still outside the giant, and during the second it is plunging-in. In simulations mn351 and ms376, three episodes of mass ejection 
were observed.  In one simulation, ms372, the mass ejection was mostly continuous. The mass ratio between first and second mass ejection varies 
from about 80\% of all unbound mass being ejected during the first mass outburst to about 36\%. The time interval between the starts of the first and last mass outbursts  varies from 1 to 9 days, and the specific kinetic energy during the first mass outbursts is larger (see, e.g., Fig S3).

As reported in Table~S1, the kinetic energy of the unbound material at infinity $E_{\rm k}^\infty$   
is $\sim10^{46}$ ergs.  
The initial energy powering the ejection $E_{\rm k}$ is larger, and is 
partly spent on overcoming the potential well just near the merged star, so $E_{\rm k}^\infty$ is not identical to $E_{\rm k}$.
In each individual simulation, $E_{\rm k}^\infty$ asymptotically approaches some 
value soon after the merger (see Fig. S3). 
We find that the total value of $E_{\rm k}^\infty$ is larger when the unbound mass is 
larger, but the specific kinetic energy is lower. 

We note that the observed V1309 Sco light-curve (see Fig.\ S1) might
well be reproduced best assuming that, as in most 
simulations, there were two or three episodes of mass ejection, where
in the first episode the unbound mass had higher ejection velocity
(Fig S4).  It is tempting to interpret this as a promising sign
for this model, and not just to the extent that adding more free parameters will
always allow a better fit. Nonetheless, we recognise that a strong conclusion
with regard to this point requires more sophisticated radiative
transfer calculations in order to produce light-curves that properly take into account
the asymmetry and structure of the ejected matter.  For the
case of ps334, the asymmetry of the photospheric surface can be seen
in Movie~S2.

\ 

\noindent{\large \bf S5. Individual objects}

\noindent\underline{\bf V838 Mon}

V838Mon's light curve is an exception as being more irregular -- its V magnitude exhibits three phases of brightening \cite{2003MNRAS.345L..25R}, with plateaus being very pronounced in bolometric luminosity (which is dominated by red spectra) during the last two peaks in V.  
This complex behavior might easily be linked to a dramatic and asymmetric start of a CEE.
One way a CEE event can start is because of the Darwin instability
(which is likely the case for V1309 Sco, as it is a system presumably
with a large mass ratio); in this case a CE is formed almost
immediately, on a dynamical timescale.  The other trigger for a CEE
event is when one of the stars starts overfilling its Roche lobe due
to evolutionary expansion (this might be the case for V838 Mon). In the latter
case, the creation of a CE does not necessarily occur on a dynamical
timescale: at the very beginning of the contact, loss of surface
layers is not expected to occur smoothly and continuously.  Mass
transfer can bring a giant out of contact, either due to a binary
expansion or due to contraction of the giant in response to sudden
mass loss, and the mass transfer unavoidably will be continued
later. This stage of evolution -- the start of the CEE in giants -- is
not well understood and is currently under debate (for a review on the
current understanding see \cite{Ivanova11}). The binary may then
proceed to a CEE on a timescale which can be estimated to be anywhere
from a year to hundreds of years \cite{podsi02,Ivanova11}. We suggest hence 
that the observed three peaks of brightening are most
likely linked to either one highly asymmetric interaction resulting
in many clumps or to several mass outbursts, similar to the case V1309
Sco discussed above, but possibly not resulting in a merger.

It is remarkable as well that it is likely that V838 Mon is a binary star now \cite{2007A&A...474..585M}, and hence may deliver several more interesting eruptive events before it completes its CEE evolution with either merger or the formation of a compact binary.

\

\noindent\underline{\bf  ILRTs from massive dusty stars}  

Some red transients have  progenitors well identified to be cool red giants surrounded by dust. 
Noticeable examples are SN 2008S and NGC 300 OT \cite{2009ApJ...699.1850B,2008ApJ...681L...9P,2009ApJ...697L..49S}.
A very good model explaining the physics of an explosion propagating through a cool expanded envelope was
developed by \cite{2011ApJ...741...37K}. A CEE outburst model as
described in our manuscript might seem -- at first sight -- not applicable.
To start with, CEE outbursts presume that an envelope is still hot enough to be ionized.

However, the reason for the explosion in that model
\cite{2011ApJ...741...37K}  is not fully established. It has been
argued that it could be an electron-capture supernova
\cite{2009ApJ...699.1850B} which occurred in an extreme asymptotic
giant branch (AGB) star. The progenitor is suggested to have been a cold
star with $T_{\rm eff}=2500K$ and $\lg_{10} (L/L_\odot)=4.6$ and 4.9, 
where the total energies radiated away in the transient stages were $3\times 10^{47}$ and$8\times 10^{47}$ erg,
for SN 2008S and NGC 300-OT respectively \cite{2011ApJ...741...37K}.

Whilst a progenitor with those properties is consistent with being a
massive AGB star, similar characteristics could also be possessed by a red giant 
during a long-term quasi-stable CEE phase known as the self-regulating
spiral-in. This situation is expected to occur if the envelope was not ejected promptly during the 
plunge-in phase; in this case the companion orbits inside the CE in a very rarefied region and the phase can last hundreds of years
 \cite{podsi02}. The stars appear to be puffed-up and somewhat cooler
 than would be predicted by standard evolutionary tracks 
\cite{iva02_mtypes}. In addition, the initial mass outburst
during the initial interaction and the plunge-in (see the discussion on
V838 Mon above) will cool down, potentially forming dust around the
system. In this case, the main orbital energy release occurs deep
inside the star,
at the bottom of the expanded envelope. For stars orbiting AGB cores,
the orbital energy release can easily be as high as a few times
$10^{48}$ ergs.

We therefore note that we can not rule out
CEE from the list of possible triggers of the explosion underlying this subclass of ILRTs, even though
the physics of the outburst is different from the WCR-determined
model,   since the outer parts of the extended envelope are cool
  enough to have recombined before the outburst.

\ 

\noindent\underline{\bf  $\eta$ Carinae}  

The Great Eruption of $\eta$ Carinae, which may also
have been triggered by a stellar merger (see, e.g.,\cite{PhP2010} and references
therein), was recently measured via light echoes, and was
susprisingly cool \cite{Rest+2012}, as would have been produced
if a recombination wave defined the photosphere.  The amount of energy
radiated was certainly too large to have been
powered purely by the recombination of a sensible
amount of material, so additional energy input would have been
required. 

\ 

\noindent{\large \bf S6. Rates}

A lower limit for the ILRT rate from observations is 
0.019 ${\rm  yr^{-1}}~L_{\rm MW}^{-1}$ \cite{Of08}, 
where $~L_{\rm MW}$ represents the luminosity of our Galaxy.  With a star formation rate of
$\approx 2 M_\odot$ per year (though rates can be from 1 to 10 $M_\odot$ per year, \cite{2011AJ....142..197C}) 
and with initial mass function from \cite{2002Sci...295...82K},
we find that, per year, roughly 0.3 stars are formed with masses large enough ($\ge 1 M_\odot$) to evolve off the MS in less than a Hubble time.  
Roughly half of stars are in binaries, and, for these initial masses, 16\% of those binaries
evolve via CEE  (with 48\% of the binaries surviving the CEE and the other 52\% experiencing a merger) 
\cite{2010ApJ...720.1752P}.  Hence the theoretically expected rate of CEEs is 
$\sim 0.024 {\rm  yr^{-1}}~L_{\rm MW}^{-1}$, in agreement with the observed lower limit.

\begin{figure}
\includegraphics[width=80mm]{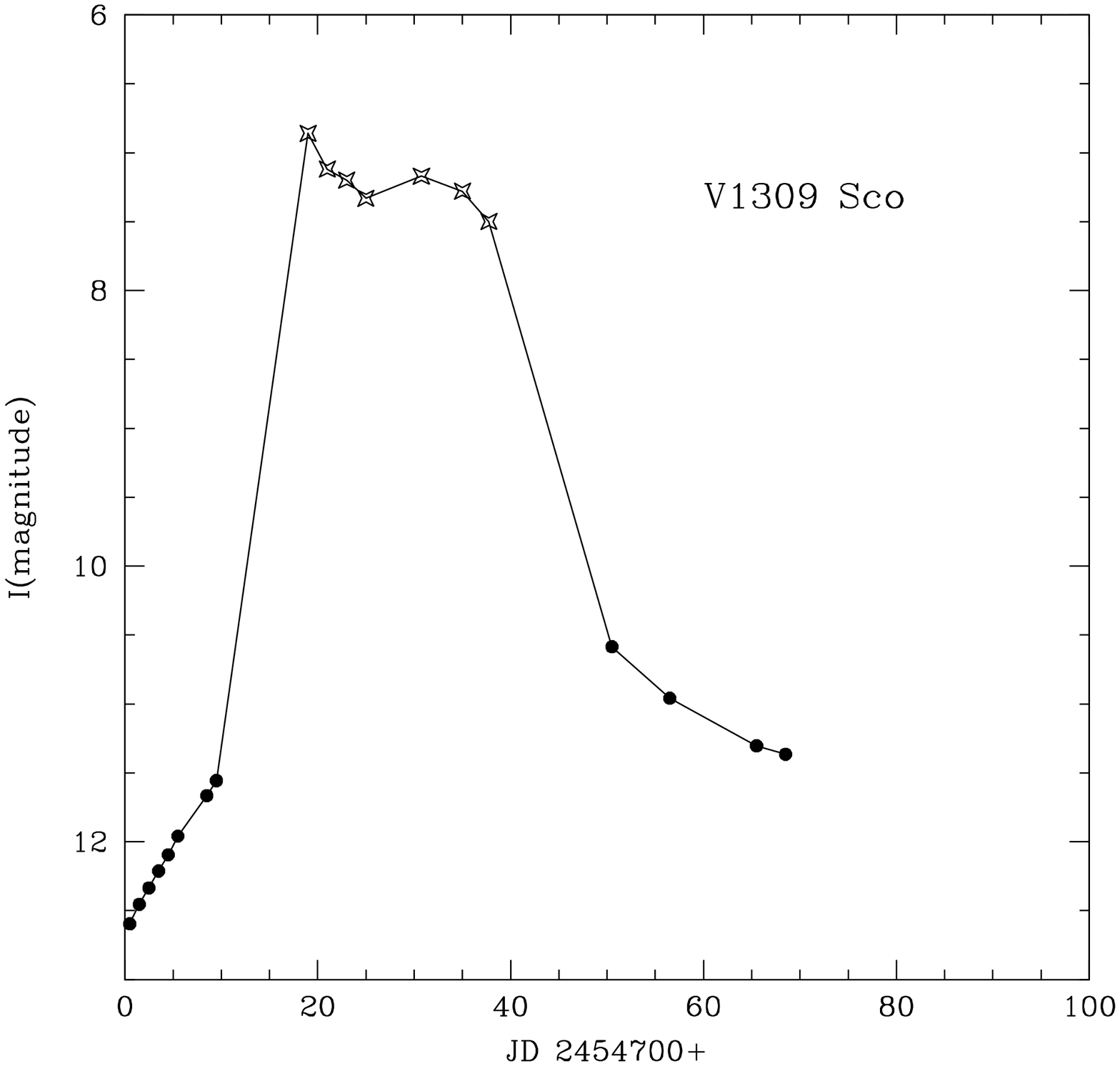}
\includegraphics[width=80mm]{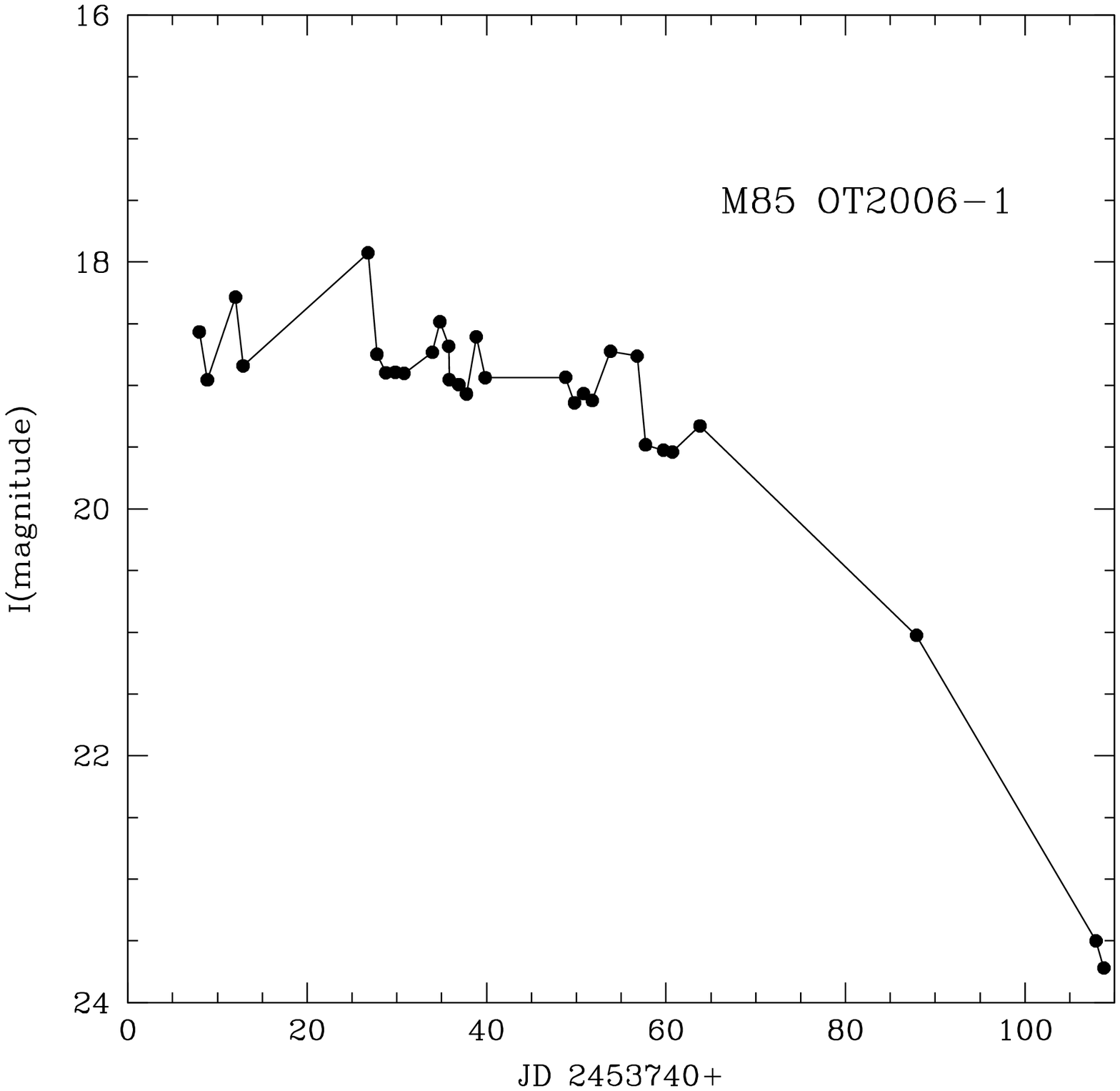}
\includegraphics[width=80mm]{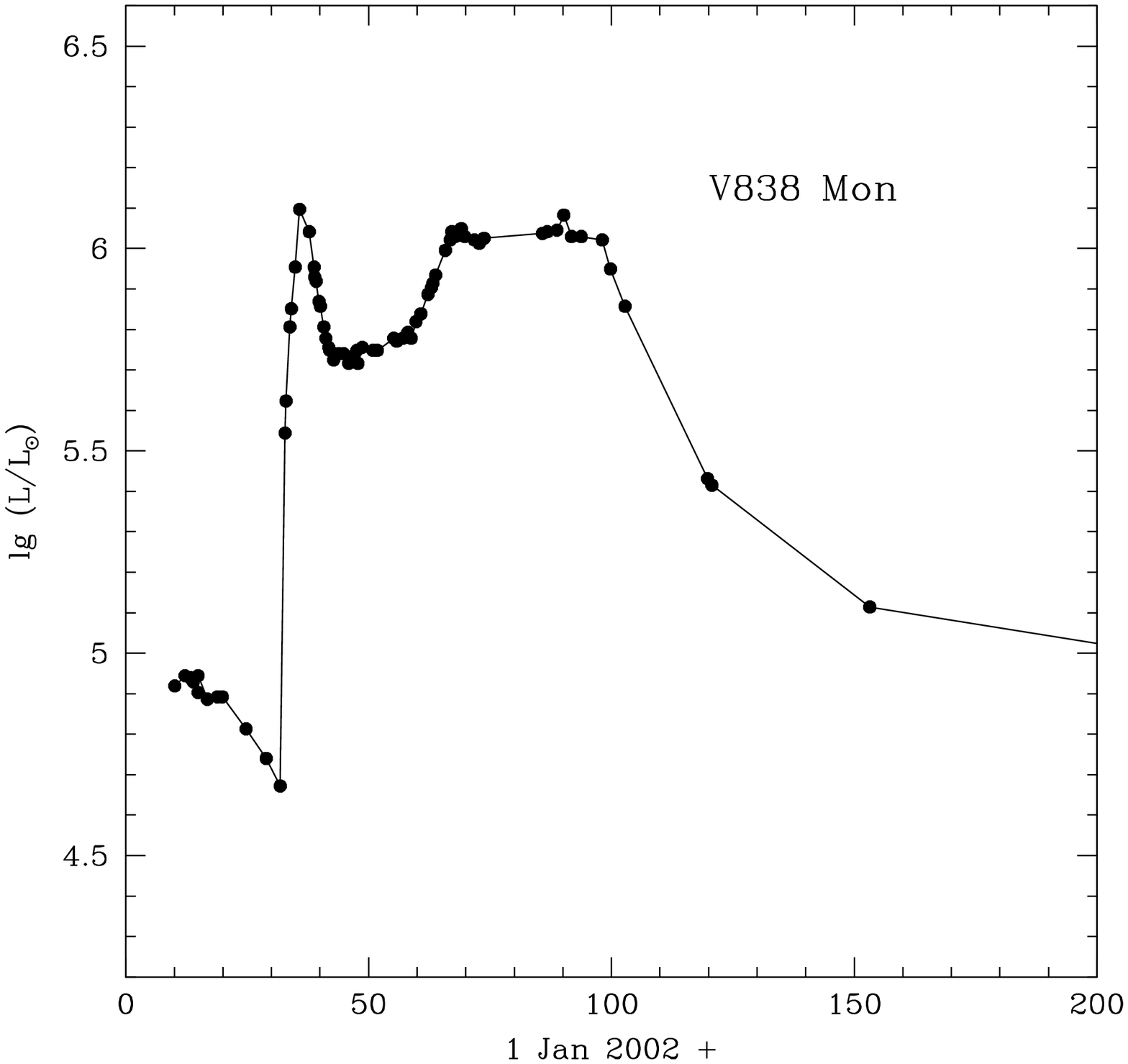}
\includegraphics[width=80mm]{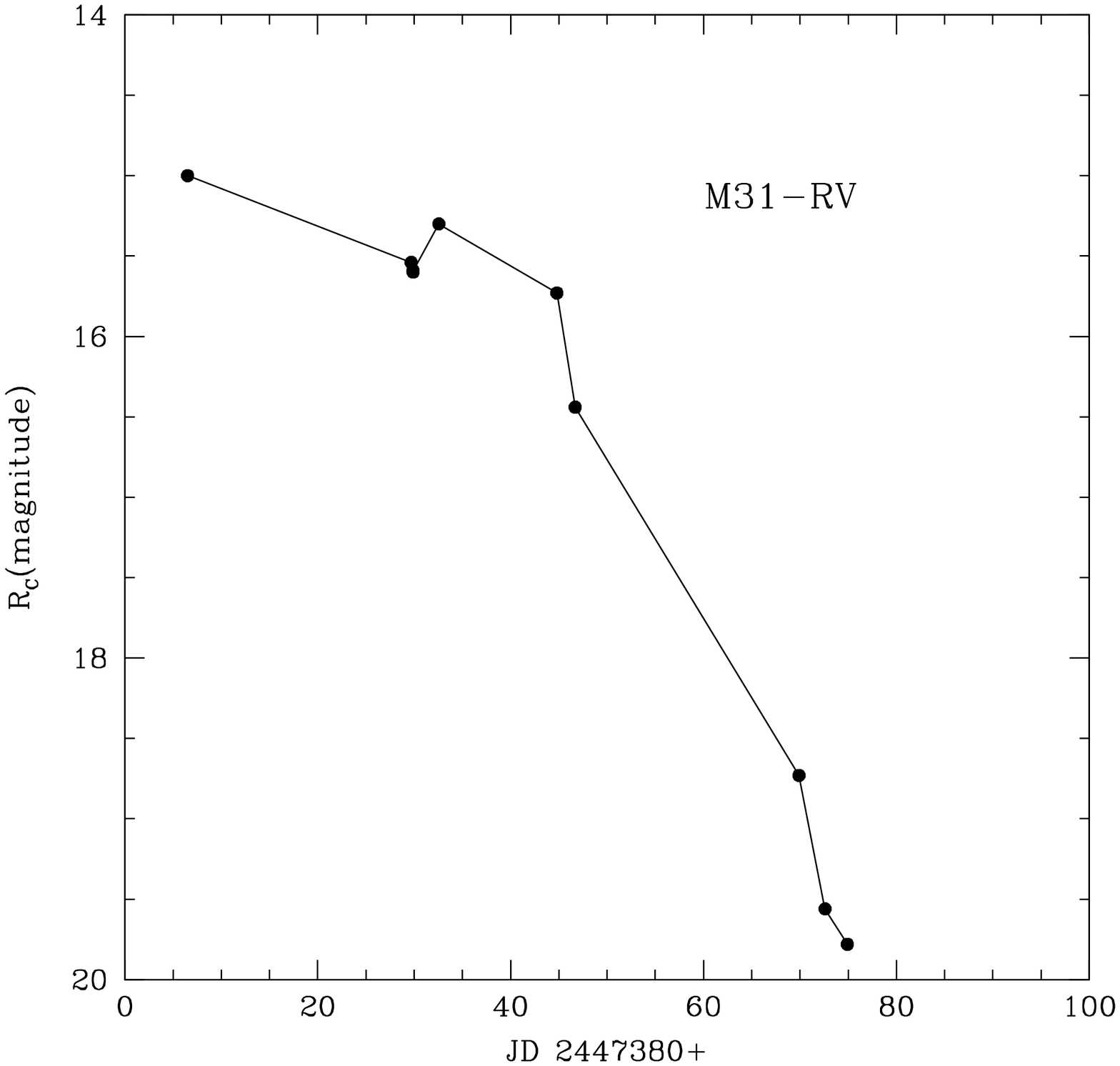}
\caption*{{\bf Fig. S1.} Observed evolution of magnitudes or luminosity
  for LRNe discussed in the text. Data was taken from
  \cite{2007Natur.449E...1P} for M85 OT2006-1, from
  \cite{2005A&A...436.1009T} for V838 Mon  and from
  \cite{2004A&A...418..869B} for M31-RV. For V1309 Sco, we used data
  from AAVSVO (available at http://www.aavso.org/),
  marked by asterisk symbols, and from OGLE-II
  \cite{2003AcA....53..291U}, marked as solid dots.
}
\end{figure}

\begin{figure}
\includegraphics[width=160mm]{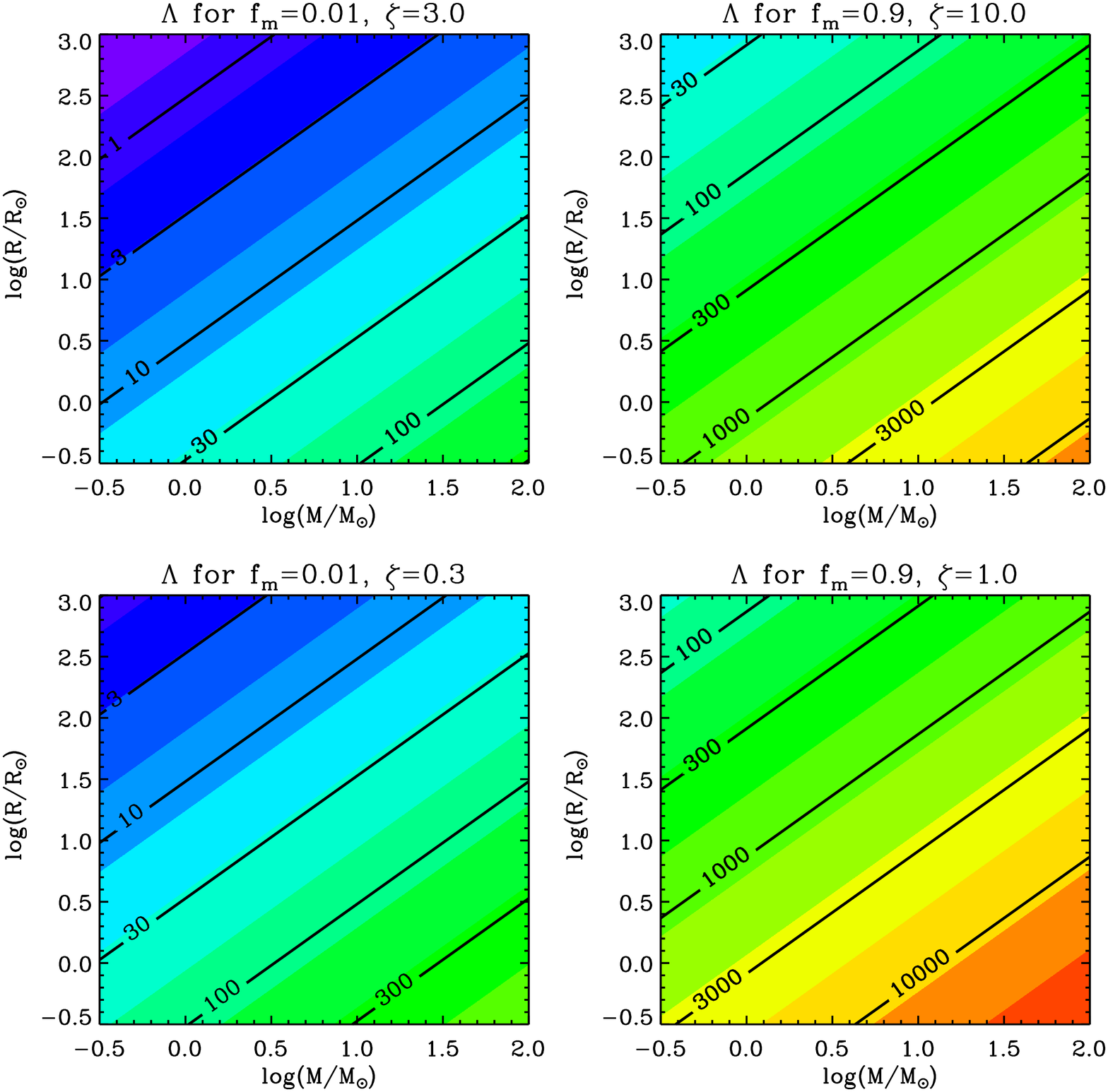}
\caption*{{\bf Fig S2.} Values of $\Lambda$, i.e.\ the parameter which indicates whether the outburst is expected to lead to a plateau, as calculated from \cite{Popov1993}. We show how these $\Lambda$-values vary across a notional pre-CEE mass-radius plane for a selection of parameter choices indicative of different potential physical situations. The two left-hand panels are estimates for mergers which eject one percent of the mass of the primary star (i.e.\ $f_{\rm m}=0.01$); the lower panel adopts $\zeta=0.3$ and the upper panel $\zeta=3$.  The two right-hand panels represent estimates for envelope ejection (i.e.\ $f_{\rm m}=0.9$); the lower panel adopts $\zeta=1$, and the upper panel $\zeta=10$ (as discussed in the text, it is reasonable to expect that $\zeta$ may typically be higher for full envelope ejection than for mergers).  Only when small amounts of matter are lost from unusually large low-mass stars does $\Lambda$ fall as low as 1. Hence we expect a plateau to be produced in the vast majority of CEE-related outburst light-curves.
}
\end{figure}

\begin{figure}
\begin{centering}
\includegraphics[width=160mm]{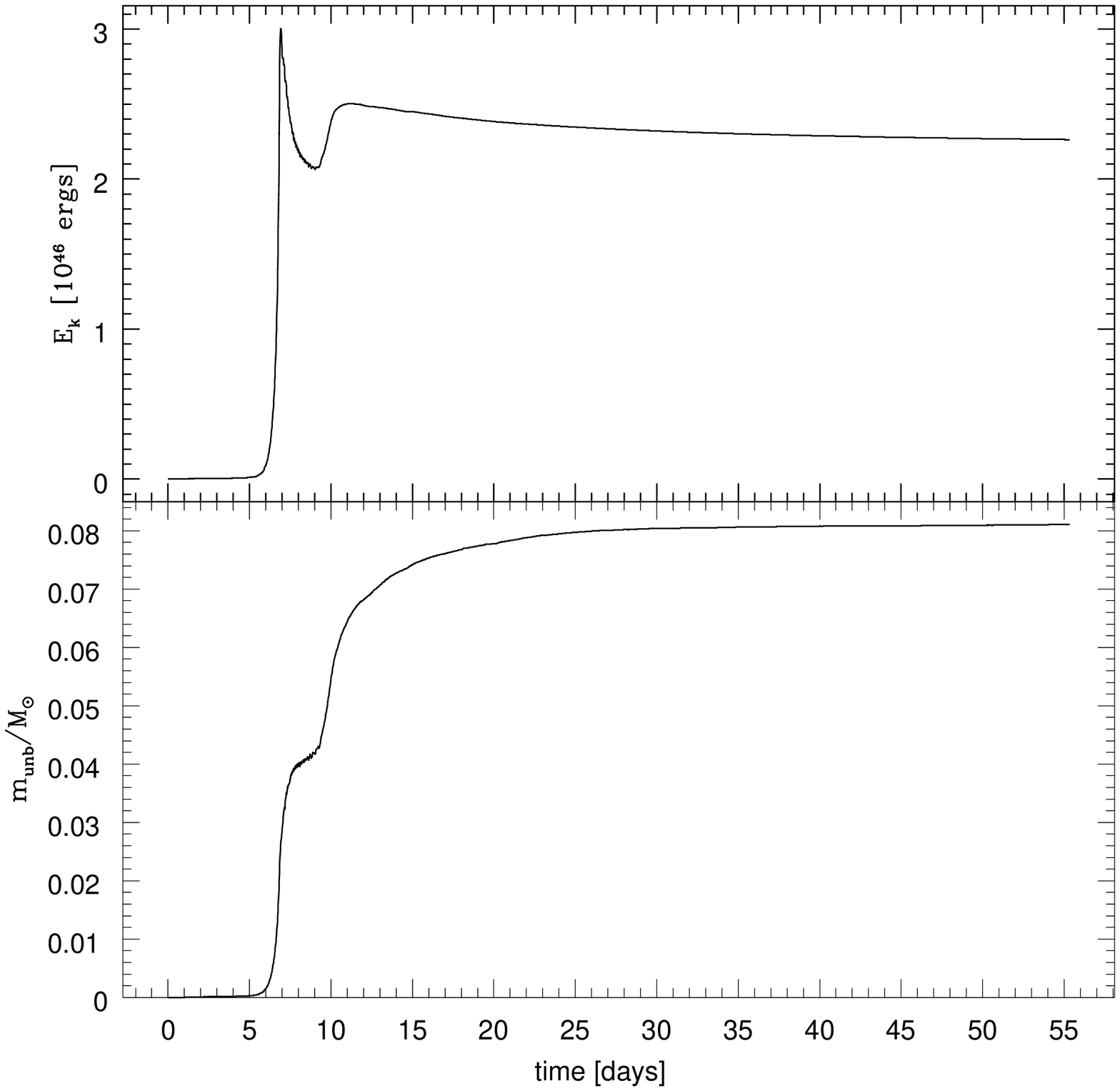}
\caption*{{\bf Fig. S3.} The time evolution of the kinetic energy $E_{\rm k}$ of the ejected material (top panel) and unbound mass $m_{\rm unb}$ (bottom panel) in simulation ps379. In the lower panel, note the two episodes of mass ejection, near times of 6 days and 10 days.}
\end{centering}
\end{figure}

\begin{figure}
\begin{centering}
\includegraphics[width=140mm]{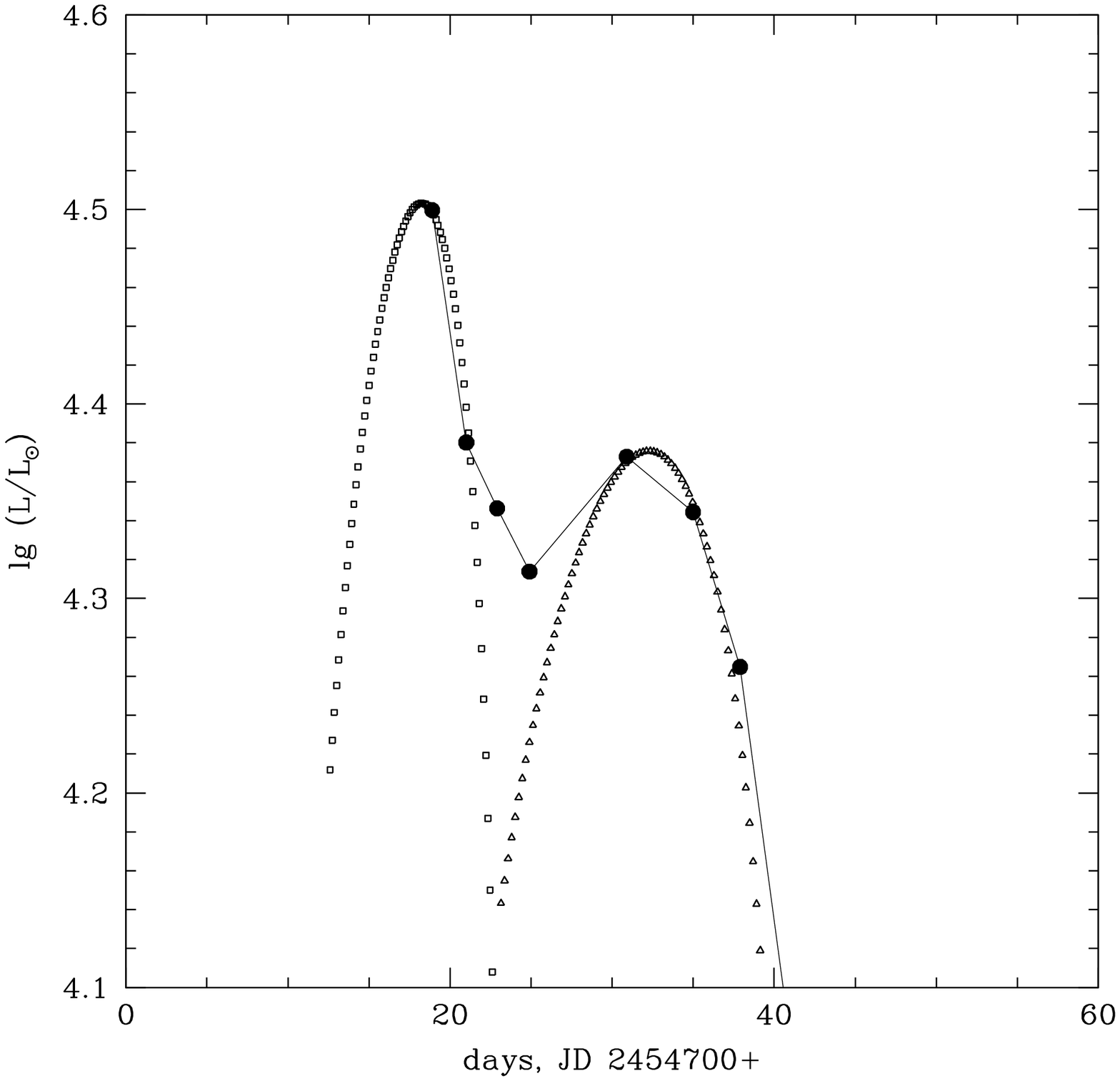}
\caption*{{\bf Fig. S4.} The observed evolution of bolometric
  luminosity during the V1309 Sco outburst \cite{Tylenda11} is
  represented by solid circles and a thin line. We compare this to  
the solutions of equations (17) and (21) of \cite{Popov1993}, 
using two consecutive mass outbursts. The light-curve
resulting from the first mass ejection is shown by squares:  $m_{\rm unb}=0.02M_\odot$, $E_{\rm k}=0.9 \times 10^{46}$ ergs,
$E_{\rm TH}^0=0.13 \times 10^{46}$ is the initial thermal energy in
this ejected layer, $E_{\rm th}=E_{\rm k}/2+ E_{\rm TH}^0$ (the
material is shocked, the standard case for Popov's model when $E_{\rm TH}^0 \ll E_{\rm k}$), the fit was done with 
$T_{\rm rec}$=5000 K, characteristic for this peak
\cite{Tylenda11}. The  light-curve
resulting from the second mass ejection is shown with triangles
$m_{\rm unb}=0.04M_\odot$, $E_{\rm k}=0.9 \times 10^{46}$ ergs,
$E_{\rm TH}^0=0.75 \times 10^{46}$ is the initial thermal energy in this
ejected layer, $E_{\rm th}=E_{\rm TH}^0$ (no shock heating in second
ejection. We adopted this for two reasons: a) this layer possesses $E_{\rm TH}^0$ 
larger than it would be suggested by Popov's approximation $E_{\rm k}/2$;
b) the second mass outburst occurs after the companion is below the ejected layer), the fit used
$T_{\rm rec}$=4000 K, characteristic for this peak
\cite{Tylenda11}. The interval between the beginnings of the two outbursts is 6.7
days.
}
\end{centering}
\end{figure}

\newpage

{
\begin{table}
\centering
\begin{tabular}{l |l l l l l l r c  |l l  l l } 
 Model &  $R_{\rm 1}^{\rm ev} $ & $R_{\rm 1}$ & $a $  & $P_{\rm orb} $ & $f_{\rm sync}$ &$f_{\rm RLOF}$ &$N_{\rm part}$ & Type  &$m_{\rm unb}$& $E_{\rm k}^\infty $ & $L_P$ & $t_{P}$ \\
\hline
 ps334  & 3.4 & 3.34 & 6.32 & 1.42 & 0.915 & 0.921 & 50000 & PM &0.042 &2.03 & 2.2e4 & 21\\
 mn351  &3.4 & 3.51 &6.32 & 1.42 &0.000& 0.968 &100000& MS  &0.039& 1.64& 2.3e4& 16 \\
 pn351 &3.4 & 3.51 &6.32 & 1.42 & 0.000 & 0.968 & 100000 & PM  &0.061 &1.63 &  1.8e4& 22 \\
 ps351 &3.4 & 3.51 &6.32 & 1.42 & 1.000 & 0.968 & 170000 & PM  & 0.065 &2.22 & 2.2e4 & 22\\
 ms376 &3.65 & 3.76 &6.55 & 1.50 & 1.000 & 0.999 & 100000 & MS  &0.040 & 1.42 & 2.0e4 & 18 \\
 ps376 &3.65 & 3.76 &6.55 & 1.50 & 1.000 & 0.999 & 100000 & PM  &0.056 &1.87 & 2.1e4&  21 \\
 ms372  &3.7 & 3.72  &6.40 & 1.45 &0.937& 1.013& 50000& MS  & 0.041&1.51& 2.1e4& 19 \\
 ps379  &3.73 & 3.79 &6.32& 1.42 &0.854 &1.045 &100000& PM& 0.081&2.25& 2.0e4& 25 \\
\end{tabular}
\caption*{ {\bf Table S1.} Results from SPH simulations. $R_{\rm 1}^{\rm
    ev}$ is the radius of the donor from the stellar code in $R_\odot$, 
$R_{\rm 1}$ is the largest spatial size of the pre-CEE donor in the SPH
 simulation after relaxation in $R_\odot$,  $a$ is the initial orbital
 separation in $R_\odot$, $P_{\rm orb}$ is the initial orbital period
 (in days) of the relaxed binary in the SPH code, $f_{\rm sync}$ gives the
 synchronization of the red giant with the orbit (0 is for no
 rotation, 1 is for fully synchronized), $N_{\rm part}$  is the number
 of particles representing the giant star, and $f_{\rm RLOF}$ is the fraction
 of the radius of the primary with respect to its Roche lobe
 radius. MS  is a main sequence star represented with 20 000 particles in mn351, 
and with 2000 particles in ms372 and ms376. PM is a point mass. The
remaining columns give results and derived quantities: 
 $ m_{\rm unb}$ is the unbound mass in $M_\odot$; $E_{\rm k}^\infty$  
is the kinetic energy at infinity in $10^{46}$ ergs;
 $L_{\rm P}$ is the plateau luminosity in $L_\odot$ calculated from equation (1) of the main text; and $t_{\rm P}$ is
 the duration of plateau phase in days, calculated from equation (2) of the main text.}
\label{table-ini}\end{table}
}
\ 

\newpage 

\ 

{\bf Movie S1}. Column densities in simulation ps334, as viewed perpendicular to the orbital plane and at times after the merger.  The merger causes an outflow of material, with the field of ejecta dropping to lower and lower densities and column densities as it expands.  At late times in the simulation, while the ejecta continues streaming outward, the bound merger product in the center of the system gradually reaches larger column densities as more material falls back to its surface. Visualizations generated using SPLASH \cite{PriceSPH}.

\

{\bf Movie S2}. Temperature cross section in the orbital plane.  Like
visualization S1, the scenario depicted  is for simulation ps334,
again viewed perpendicular to the orbital plane at times after the
merger.  Only temperatures between 4000 K and 5000 K are shown,
corresponding roughly to the recombination temperature and hence
roughly to the location of the photosphere.  Note that between $\sim$
165 and 185 days, lines of sight from an observer looking along the
orbital plane do not fully penetrate down to the innermost closed
4500K surface: this enlarged photosphere is the main cause of the a
plateau phase.  Near $\sim$ 185 days, the shroud dissipates and the
inner bound merger product would be revealed, corresponding to the
decline stage after the plateau.  Visualizations generated using SPLASH \cite{PriceSPH}.

\

\noindent{\bf Acknowledgments:} N.I. acknowledges support from the
Natural Sciences and Engineering Research Council of
Canada and the Canada Research Chairs Program. S.J. acknowledges support from
the Chinese Academy of Sciences and National Natural Science
Foundation of China. J.L.A.N. acknowledges support from
CONACyT. J.C.L. thanks R. Scruggs for useful discussions. This
research has been enabled by the use of computing resources
provided by WestGrid and Compute/Calcul Canada as well
as the Extreme Science and Engineering Discovery
Environment (supported by NSF grant OCI-1053575).


\newpage

\begin{thebibliography}{10}

\bibitem{Ivanova11}
N.~{Ivanova}, {\it ASP Conference Series} {\bf 447} , 91 (2011)

\bibitem{som}
Materials, methods are available as supplementary materials~on Science~Online .

\bibitem{1991SvAL...17..210C}
N.~N. {Chugai}, {\it Soviet Astronomy Letters\/} {\bf 17}, 210 (1991).

\bibitem{Popov1993}
D.~V. {Popov}, {\it \apj\/} {\bf 414}, 712 (1993).

\bibitem{Kasen09}
D.~{Kasen}, S.~E. {Woosley}, {\it \apj\/} {\bf 703}, 2205 (2009).

\bibitem{2010ApJ...714..155K}
D.~{Kasen}, E.~{Ramirez-Ruiz}, {\it \apj\/} {\bf 714}, 155 (2010).

\bibitem{1993PASP..105.1373I}
I.~{Iben}, Jr., M.~{Livio}, {\it \pasp\/} {\bf 105}, 1373 (1993).

\bibitem{Han+1994}
Z.~{Han}, P.~{Podsiadlowski}, P.~P. {Eggleton}, {\it \mnras\/} {\bf 270}, 121
  (1994).

\bibitem{Web08}
R.~F. {Webbink} (2008), vol. 352 of {\it ASSL\/}, p. 233.


\bibitem{1990ApJ...353L..35M}
J.~{Mould}, {\it et~al.\/}, {\it \apjl\/} {\bf 353}, L35 (1990).

\bibitem{1999AJ....118.1034M}
P.~{Martini}, {\it et~al.\/}, {\it \aj\/} {\bf 118}, 1034 (1999).

\bibitem{Bond+2003}
H.~E. {Bond}, {\it et~al.\/}, {\it \nat\/} {\bf 422}, 405 (2003).

\bibitem{2007Natur.447..458K}
S.~R. {Kulkarni}, {\it et~al.\/}, {\it \nat\/} {\bf 447}, 458 (2007).

\bibitem{2007Natur.449E...1P}
A.~{Pastorello}, {\it et~al.\/}, {\it \nat\/} {\bf 449} (2007).

\bibitem{Bond+2009}
H.~E. {Bond}, {\it et~al.\/}, {\it \apjl\/} {\bf 695}, L154 (2009).

\bibitem{2011ApJ...737...17B}
H.~E. {Bond}, {\it \apj\/} {\bf 737}, 17 (2011).


\bibitem{Kasliwal+2011}
M.~M. {Kasliwal}, {\it et~al.\/}, {\it \apj\/} {\bf 730}, 134 (2011).

\bibitem{2009ApJ...699.1850B}
E.~{Berger}, {\it et~al.\/}, {\it \apj\/} {\bf 699}, 1850 (2009).

\bibitem{Tylenda11}
R.~{Tylenda}, {\it et~al.\/}, {\it \aap\/} {\bf 528}, A114 (2011).


\bibitem{Of08}
E.~O. {Ofek}, {\it et~al.\/}, {\it \apj\/} {\bf 674}, 447 (2008).

\bibitem{Rau07}
A.~{Rau}, S.~R. {Kulkarni}, E.~O. {Ofek}, L.~{Yan}, {\it \apj\/} {\bf 659},
  1536 (2007).

\bibitem{Stepien11}
K.~{St{\c e}pie{\'n}}, {\it \aap\/} {\bf 531}, A18 (2011).

\bibitem{Rau09}
A.~{Rau}, {\it et~al.\/}, {\it \pasp\/} {\bf 121}, 1334 (2009).

\bibitem{Hurley+2000}
J.~R. {Hurley}, O.~R. {Pols}, C.~A. {Tout}, {\it \mnras\/} {\bf 315}, 543
  (2000).


\bibitem{2012AAS...21943609B}
H.~E. {Bond}, {\it et~al.\/}, {\it American Astronomical Society Meeting
  Abstracts\/} (2012), vol. 219 of {\it American Astronomical Society Meeting
  Abstracts\/}, p. 436.09.

\bibitem{2012ApJ...746..100S}
N.~{Soker}, A.~{Kashi}, {\it \apj\/} {\bf 746}, 100 (2012).

\bibitem{2011MNRAS.416.1965B}
E.~{Bear}, A.~{Kashi}, N.~{Soker}, {\it \mnras\/} {\bf 416}, 1965 (2011).

\bibitem{1992ApJ...389..369I}
I.~{Iben}, Jr., A.~V. {Tutukov}, {\it \apj\/} {\bf 389}, 369 (1992).

\bibitem{2009ApJ...705.1364T}
T.~A. {Thompson}, {\it et~al.\/}, {\it \apj\/} {\bf 705}, 1364 (2009).

\bibitem{2011ApJ...741...37K}
C.~S. {Kochanek}, {\it \apj\/} {\bf 741}, 37 (2011).

\bibitem{2006MNRAS.373..733S}
N.~{Soker}, R.~{Tylenda}, {\it \mnras\/} {\bf 373}, 733 (2006).

\bibitem{2005A&A...439..651T}
R.~{Tylenda}, L.~A. {Crause}, S.~K. {G{\'o}rny}, M.~R. {Schmidt}, {\it \aap\/}
  {\bf 439}, 651 (2005).

\bibitem{2008ApJ...681L...9P}
J.~L. {Prieto}, {\it et~al.\/}, {\it \apjl\/} {\bf 681}, L9 (2008).

\bibitem{2009ApJ...697L..49S}
N.~{Smith}, {\it et~al.\/}, {\it \apjl\/} {\bf 697}, L49 (2009).

\bibitem{2007ASPC..363..189S}
M.~H. {Siegel}, H.~E. {Bond}, {\it The Nature of V838 Mon and its Light
  Echo\/}, R.~L.~M. {Corradi}, U.~{Munari}, eds. (2007), vol. 363 of {\it
  Astronomical Society of the Pacific Conference Series\/}, p. 189.

\bibitem{2005A&A...436.1009T}
R.~{Tylenda}, {\it \aap\/} {\bf 436}, 1009 (2005).

\bibitem{2004A&A...418..869B}
F.~{Boschi}, U.~{Munari}, {\it \aap\/} {\bf 418}, 869 (2004).

\bibitem{2003AcA....53..291U}
A.~{Udalski}, {\it Acta Astronomica\/} {\bf 53}, 291 (2003).

\bibitem{2003ApJ...582L.105S}
N.~{Soker}, R.~{Tylenda}, {\it \apjl\/} {\bf 582}, L105 (2003).

\bibitem{Passy12}
J.-C. {Passy}, {\it et~al.\/}, {\it \apj\/} {\bf 744}, 52 (2012).

\bibitem{Ricker12}
P.~M. {Ricker}, R.~E. {Taam}, {\it \apj\/} {\bf 746}, 74 (2012).

\bibitem{TylendaSoker05}
R.~{Tylenda}, N.~{Soker}, R.~{Szczerba}, {\it \aap\/} {\bf 441}, 1099 (2005).


\bibitem{crause03}
Crause, L.~A., Lawson, 
W.~A., Kilkenny, D., et al.\ 2003, \mnras, 341, 785 


\bibitem{2003MNRAS.345L..25R}
A.~{Retter}, A.~{Marom}, {\it \mnras\/} {\bf 345}, L25 (2003).

\bibitem{2003ApJ...588..486W}
J.~P. {Wisniewski}, {\it et~al.\/}, {\it \apj\/} {\bf 588}, 486 (2003).

\bibitem{ImsNad1965}
V.~S. {Imshennik}, D.~K. {Nadyozhin}, {\it \sovast\/} {\bf 8}, 664 (1965).

\bibitem{Grassberg+1971}
E.~K. {Grassberg}, V.~S. {Imshennik}, D.~K. {Nadyozhin}, {\it \apss\/} {\bf
  10}, 28 (1971).

\bibitem{GraNad1976}
E.~K. {Grassberg}, D.~K. {Nadyozhin}, {\it \apss\/} {\bf 44}, 409 (1976).

\bibitem{Eastman+1994}
R.~G. {Eastman}, S.~E. {Woosley}, T.~A. {Weaver}, P.~A. {Pinto}, {\it \apj\/}
  {\bf 430}, 300 (1994).

\bibitem{ZKR1958a}
Y.~B. {Zel'Dovich}, A.~{Kompaneets}, Y.~P. {Raizer}, {\it JETP\/} {\bf 7}, 882
  (1958).

\bibitem{ZKR1958b}
Y.~B. {Zel'Dovich}, A.~{Kompaneets}, Y.~P. {Raizer}, {\it JETP\/} {\bf 7}, 1001
  (1958).

\bibitem{Bethe1964}
H.~{Bethe}, {\it {Los Alamos Report}\/} {\bf {LA-3064}}, {44} ({1964}).

\bibitem{Kochanek+2012}
C.~S.~{Kochanek}, D.~M.~{Szczygie{\l}}, K.~Z.~{Stanek}, {\it \apj}
{\bf 758}, 142 (2012). 

\bibitem{1985SvAL...11..145L}
I.~Y. {Litvinova}, D.~K. {Nadezhin}, {\it Soviet Astronomy Letters\/} {\bf 11},
  145 (1985).

\bibitem{2011ApJ...743...49L}
A.~J. {Loveridge}, M.~V. {van der Sluys}, V.~{Kalogera}, {\it \apj\/} {\bf
  743}, 49 (2011).

\bibitem{Egg71}
P.~P. {Eggleton}, {\it \mnras\/} {\bf 151}, 351 (1971).

\bibitem{Egg72}
P.~P. {Eggleton}, {\it \mnras\/} {\bf 156}, 361 (1972).

\bibitem{Egg73}
P.~P. {Eggleton}, {\it \mnras\/} {\bf 163}, 279 (1973).

\bibitem{Eggetal73}
P.~P. {Eggleton}, J.~{Faulkner}, B.~P. {Flannery}, {\it \aap\/} {\bf 23}, 325
  (1973).

\bibitem{Pols95}
O.~R. {Pols}, C.~A. {Tout}, P.~P. {Eggleton}, Z.~{Han}, {\it \mnras\/} {\bf
  274}, 964 (1995).

\bibitem{Glebbeek2008}
E.~{Glebbeek}, O.~R. {Pols}, J.~R. {Hurley}, {\it \aap\/} {\bf 488}, 1007
  (2008).

\bibitem{Lombardi06}
J.~C. {Lombardi}, Jr., {\it et~al.\/}, {\it \apj\/} {\bf 640}, 441 (2006).

\bibitem{Gaburov10}
E.~{Gaburov}, J.~C. {Lombardi}, Jr., S.~{Portegies Zwart}, {\it \mnras\/} {\bf
  402}, 105 (2010).

\bibitem{Lombardi11}
J.~C. {Lombardi}, Jr., {\it et~al.\/}, {\it \apj\/} {\bf 737}, 49 (2011).

\bibitem{Renvoize+2002}
V.~{Renvoiz{\'e}}, I.~{Baraffe}, U.~{Kolb}, H.~{Ritter}, {\it \aap\/} {\bf
  389}, 485 (2002).

\bibitem{podsi02}
Podsiadlowski, P.\ 2001, 
Evolution of Binary and Multiple Star Systems, 229, 239 

\bibitem{2007A&A...474..585M}
U.~{Munari}, {\it et~al.\/}, {\it \aap\/} {\bf 474}, 585 (2007).

\bibitem{iva02_mtypes}
N.~{Ivanova}, P.~{Podsiadlowski}, {\it Exotic Stars as Challenges to
  Evolution\/}, {C.~A.~Tout \& W.~van Hamme}, ed. (2002), vol. 279 of {\it
  Astronomical Society of the Pacific Conference Series\/}, pp. 245--+.

\bibitem{PhP2010}
P.~{Podsiadlowski}, {\it \nar\/} {\bf 54}, 39 (2010).

\bibitem{Rest+2012}
A.~{Rest}, {\it et~al.\/}, {\it \nat\/} {\bf 482}, 375 (2012).

\bibitem{2011AJ....142..197C}
L.~{Chomiuk}, M.~S. {Povich}, {\it \aj\/} {\bf 142}, 197 (2011).

\bibitem{2002Sci...295...82K}
P.~{Kroupa}, {\it Science\/} {\bf 295}, 82 (2002).

\bibitem{2010ApJ...720.1752P}
M.~{Politano}, M.~{van der Sluys}, R.~E. {Taam}, B.~{Willems}, {\it \apj\/}
  {\bf 720}, 1752 (2010).

\bibitem{PriceSPH}
D.J.~Price, {\it PASA},  {\bf 24}, 159 (2007)

\end{thebibliography}
\end{document}